\documentclass[12pt]{article}

\usepackage{amscd, amsmath}
\usepackage{amsthm, amssymb}
\newcommand{\CC}{\mathbb{C}}
\newcommand{\GME}{\Gamma (E)}
\newcommand{\AAA}{\mathcal A}
\newcommand{\OO}{\mathcal O}
\newcommand{\OOO}{\mathbb O}
\newcommand{\he}{\hat{e}}
\newcommand{\hphi}{\hat{\phi}}
\newcommand{\db}{\bar{\partial}}
\newcommand{\dd}{{\rm d}}

\newcommand{\g}{\mathfrak g}
\newcommand{\Gg}{\mathfrak G}
\newcommand{\GG}{\mathcal G}
\newcommand{\UU}{U}
\newcommand{\VV}{\mathcal W}
\newcommand{\VVV}{\mathcal V}
\newcommand{\ver}{{\rm Vert}}

\newtheorem{thm}{Theorem}[section]
\newtheorem{prop}[thm]{Proposition}

\newtheorem*{thm*}{Theorem}
\newtheorem*{cor*}{Corollary}

\theoremstyle{remark}
\newtheorem*{rem}{{\bf Remark}}

\theoremstyle{remark}
\newtheorem*{prop*}{{\bf Proposition}}

\theoremstyle{remark}
\newtheorem*{defn}{{\bf Definition}}

\theoremstyle{remark}
\newtheorem*{lem*}{{\bf Lemma}}

\theoremstyle{remark}

\theoremstyle{remark}

\begin{document}

\begin{titlepage}
\title{
\vskip -70pt
\begin{flushright}
{\normalsize \ DAMTP-2005-18}\\
\end{flushright}
\vskip 45pt
{\bf A topological gauged sigma-model}
}
\vspace{3cm}

\author{{J. M. Baptista} \thanks{ e-mail address:
    J.M.Baptista@damtp.cam.ac.uk }  \\
{\normalsize {\sl Department of Applied Mathematics and Theoretical
    Physics} \thanks{ address: Wilberforce Road, Cambridge
    CB3 0WA, England }
} \\
{\normalsize {\sl University of Cambridge}} 
}

\date{February 2005}

\maketitle

\thispagestyle{empty}
\vspace{2cm}
\vskip 20pt
{\centerline{{\large \bf{Abstract}}}}
\vspace{.35cm}
We describe a topological field theory that studies the moduli space
of solutions of the symplectic vortex equations. It contains as
special cases the topological sigma-model and topological Yang-Mills
over K\"ahler surfaces. The correlation functions of the theory are
closely related to the recently introduced Hamiltonian Gromov-Witten
invariants. 
\end{titlepage}

\tableofcontents

\section{Introduction}

Topological field theories are one of the cornerstones of the modern
relations between theoretical physics and mathematics. Their originality
stems from the fact that they employ
methods of quantum field theory to study problems in geometry; most
notably, they use path-integrals to obtain invariants of manifolds. 
The first explicit examples of these theories were topological Yang-Mills and
the topological sigma-model, introduced by Witten in \cite{W1} and
\cite{W2}. Topological Yang-Mills studies the moduli space of
instantons over a four-manifold. Its correlation functions are then closely
related to the Donaldson invariants. The topological sigma-model
studies pseudo-holomorphic curves on an almost K\"ahler manifold, and
its correlation functions are essentially the Gromov-Witten
invariants. After these initial examples several other topological
theories were introduced, for instance 2D topological gravity,
Chern-Simons theory and topological string theory, each studying
different moduli spaces and invariants. This, however, is very
well known story.

The aim of this paper is to define a topological field theory that
studies the moduli space of solutions of the symplectic vortex
equations. Its correlation functions, as we will see, are then closely
related to the so-called Hamiltonian Gromov-Witten invariants. Both
the vortex equations and the latter invariants were recently
introduced in the mathematics literature by Mundet i Riera in
\cite{MiR}, and by Cieliebak, Gaio and Salamon in \cite{C-G-S}. They
have been further studied in 
\cite{C-G-M-S, G-S, M-T}. Here we want to give a topological field
theory version of the subject.

$\ $

The setting for the theory is a non-linear gauged sigma-model with
K\"ahler domain $M$ and almost K\"ahler target $X$. The manifold $X$
should also be equipped with a hamiltonian and holomorphic action of
the gauge group $G$. The fields of the theory are then the maps $\phi$
between $M$ and $X$ and the $G$-connections $A$ over $M$. The energy
functional is defined as 
\[
\mathcal{E}(A,\phi )\ =\  \int_M \;   \, 
\| F_A \|^2  +  \| \dd^A   \phi \|^2 +   \| \mu \circ \phi \|^2 \ ,  
\]
where $F_A$ is the curvature of $A$, $\dd^A \phi$ is a covariant
derivative, and $\mu$ is a moment map for the $G$-action on
$X$. Notice that we are not simply gauging the usual sigma-model,
since a Maxwell term and a very important potential term are also
present. By a Bogomolny argument it can be shown that
the energy is minimized by the solutions of 
\begin{subequations} 
\begin{align*}
 &  \db^A \phi \ =\  0        \\
 &  \Lambda F_A + \ \mu\circ\phi \ = \ 0   \\
 &  F_A^{0,2}\ =\ 0  \ \ ,   
\end{align*}
\end{subequations}
which are the general vortex equations. All the notation is explained
in detail in the next section. Notice that when the group $G$ is
trivial the energy functional and the vortex equations reduce to the
usual sigma-model. When the manifold $M$ is four-dimensional and $X$
is a point, we obtain instead 
the Yang-Mills energy and the equations $\Lambda F_A =0$ and
$F_A^{0,2} = 0$, which are just the anti-self-duality equations in
disguise. Thus the topological field theory that we want to define
will contain as special cases the topological sigma-model (or more
precisely the A-model) and topological Yang-Mills over K\"ahler
surfaces. It should also contain the topological gauged linear
sigma-model constructed by Witten in \cite{W3}; this corresponds to
taking $M$ a Riemann surface, $X$ a complex vector space, and $G$ a
unitary group.

All throughout the paper we approach the topological theory from the
point of view of infinite-dimensional differential geometry and
equivariant cohomology.
%% It would be interesting to see in which instances the same theory
%% can be obtained by twisting a supersymmetric theory, as was done in
%% \cite{W3}. 
This geometrical point of view was pioneered
in \cite{A-J} for topological Yang-Mills, and was subsequently applied
to many other field theories. It is reviewed for example in
\cite{B-B-R-T, Bl, C-M-R}. Also, here the manifold $X$ is always
assumed to be K\"ahler, although the formalism could presumably be
extended to the almost K\"ahler case.
 
$\ $

We will now give a brief description of the content of each
section. Sections 2 and 3 are introductory. In the first one we review
the basic facts about the Yang-Mills-Higgs functional and the general
vortex equations. In the second one we introduce the geometrical
approach to the space of fields of the gauged sigma-model. This consists
of an informal presentation of the infinite-dimensional manifold of
fields, its tangent space and 1-forms, the action of the group of
gauge transformations, and the associated Cartan model for
equivariant cohomology. Here we try to give a careful exposition and
introduce some necessary notation, but all the material is standard.

In Section 4 we contruct the topological Lagrangian for the gauged
sigma-model. As is usually the case for cohomological field theories,
this Lagrangian can be obtained from purely geometrical
considerations. Roughly speaking, it comes directly from the
Mathai-Quillen representative of the Thom class of a certain
infinite-dimensional vector bundle over the moduli space of
fields \cite{A-J, B-B-R-T, Bl, C-M-R}. In
Appendix A we explain how the gauged sigma-model can be fitted into
this geometrical approach. The results obtained there can then be fed
into the standard procedures of \cite{A-J, B-B-R-T, Bl, C-M-R} in order to
justify many of the apparently arbitrary choices in Section 4. The
approach of Section 4, by itself, is a very ``nuts and bolts'' one,
more along the lines of the original constructions in \cite{W1, W2}.

In Section 5 we define the natural observables of the theory.
%% whose correlation functions essentialy correspond to the
%% Hamiltonian Gromov-Witten invariants.
In the geometric picture, these observables are just a set of 
closed elements of the equivariant complex of the space of
fields. Section 6 is then spent explaining in detail the relation
between these ``quantum'' observables and the more traditional ways of
defining invariants, namely the so-called universal
contructions. All the work of this section (and of most of
the paper, by the way) is just a matter of suitably combining and
generalizing well known constructions from topological Yang-Mills and
the topological sigma-model. 

In the first part of Section 7 we apply standard localization to write down the
correlation functions as integrals of differential forms over the
moduli space of vortex solutions. Using the results
of Section 6, this finally allows us to
compare the correlation functions of the topological theory with the
Hamiltonian Gromov-Witten invariants defined in \cite{C-G-M-S}. In
this reference the invariants have been rigorously defined for $M$ a Riemann
surface and a suitable class of $G$-manifolds $X$.
The last two subsections are
then mostly informative: in 7.2 we discuss the moduli space of vortex
solutions in the case where $M$ is a Riemann surface;  in $7.3$ we
comment on some features of the invariants, namely the wall-crossing
phenomena and the adiabatic limit of the vortex equations. Appendix B
contains the proof of a proposition stated in 7.2 about the vortex
moduli space for torus actions.

\section{The vortex equations}

In this section we will go through a quick review of the gauged
sigma-model that admits vortex equations. For more details see for example
the original references \cite{MiR, C-G-S} or the first section of \cite{Ba}.

$\ $

The data we need to define the sigma-model are the following.
\begin{itemize}
\item[$\bullet$] Two K\"ahler manifolds $M$ and $X$, with respective
K\"ahler forms $\omega_M$ and $\omega_X$.
\item[$\bullet$] A connected compact Lie group $G$ with Lie algebra $\g$,
and an Ad-invariant positive-definite inner product $\kappa(\cdot ,
\cdot)$  on $\g$.
\item[$\bullet$] An effective, hamiltonian, left action $\rho$ of $G$ on $X$
such that, for every $g\in G$, the transformations $\rho_g : X \rightarrow
X$ are holomorphic, and a moment map for this action $\mu : X \rightarrow 
\g^\ast$.
\item[$\bullet$] A principal $G$-bundle $\pi_P : P \rightarrow M$.
\end{itemize}
We remark that, in the fullest generality, the complex structure on
$X$ need not be assumed integrable, but we will assume that here.
Using the elements above one can define the associated bundle $E =
P\times_\rho X$, which is a bundle over $M$ with typical fiber $X$. It is
defined as the quotient of $P\times X$ by the equivalence relation
$(p\cdot g ,q) \sim (p, g \cdot q)$, for all  $g\in G$.  The bundle    
projection $\pi_E : E \rightarrow M$ is determined by $\pi_E ([p,q]) \
=\ \pi_P (p)$, where $[p,q]$ denotes the equivalence class in $E$ of
the point $(p,q)$ in $P \times X$.

\vspace{.2cm}

\begin{defn}
The convention used here is that a moment map for the action $\rho $
of $G$ on $(X, \omega_X )$ is a map $\mu : X \rightarrow \g^\ast $ such
that
\begin{itemize}
\item[(i)]$\dd\, \langle \mu , \xi \rangle = \iota_{\hat{\xi}}\,
  \omega_X $ in $\Omega^1 
  (X)$ for all $\xi \in \g $, where $\hat{\xi}$ is the vector field on $X$
  defined by the flow $t \mapsto \rho_{\exp (t\xi )}$.
\item[(ii)] $\rho_g^\ast \, \mu = {\rm Ad}_g^\ast \circ \mu $ for all $g
  \in G$, where ${\rm Ad}_g^\ast $ is the coadjoint representation of
  $G$ on $\g^\ast$.
\end{itemize}
If a moment map $\mu$ exists, it is not in general unique, but all the
other moment maps are of the form $\mu + c$, where $c \in [\g ,\g ]^0
\subset \g^\ast $ is a constant in the annihilator of $[\g, \g]
$. Recall also that under the identification $\g^\ast \simeq \g$
provided by an ${\rm Ad}$-invariant inner product on $\g$, the annihilator
$[\g, \g ]^0 $ is taken to the centre of $\g$. 
\end{defn}

\vspace{.2cm}

The fields of the theory are a connection $A$ on the principal bundle $P$
and a smooth section $\phi $ of $E$. Calling $\AAA$ the space of such   
connections and $\Gamma (E)$ the space of such sections, we define the
energy functional $\mathcal{E} : \AAA \times \Gamma (E) \rightarrow    
{\mathbb R}_0^+$ of the sigma-model by
\begin{gather}
\mathcal{E}(A,\phi )\ =\ \frac{1}{2} \int_M \;   \frac{1}{e^2} \, 
\| F_A \|^2  + 
\| \dd^A   \phi \|^2 + e^2 \, \| \mu \circ \phi \|^2   
 \; ,\qquad  e\in {\mathbb R}^+ .
\label{2.1}
\end{gather}
In this formula $F_A$ is the curvature of the connection $A$, and
$\dd^A \phi$ is the covariant derivative of $\phi$ induced by $A$. The
norms are defined in the natural way, using the metrics on $M$, $X$
and $\g$. The last term is well defined because of the
$G$-equivariance of the moment map and the ${\rm Ad}_G$-invariance of
the inner product $\kappa$.

$\ $

For later convenience we will record here the local (i.e
trivialization-dependent) formulae for $\dd^A \phi$ and $\db^A
\phi$. Let $s : \UU \rightarrow P$ be a local section of $P$ over a
domain $\UU$ in $M$. Since $E=P\times_\rho X$ is an associated bundle,
this determines a trivialization of $E|_\UU$ by
\begin{gather}
\UU \times X \ \simeq \ E|_\UU \ , \qquad (x,q) \ \simeq \
 [ s(x), q] \ .
\label{2.2}
\end{gather}
With respect to these trivializations a section $\phi$ of $E$ can be
locally identified with a map $\hat{\phi} : \UU \rightarrow X$, and a
connection $A$ on $P$ can be identified with the connection form
$s^\ast A = A_\mu \; \dd x^\mu   \in \Omega^1 (\UU ; \g)$. Then the covariant
derivative $\dd^A \phi$, which is a section of the bundle $T^\ast M
\otimes\, \phi^\ast \ker \dd\, \pi_E \rightarrow M $, is locally given by 
\begin{equation}
\dd^A \phi \ =\ \dd \, \hat{\phi} \ +\ s^\ast A^a \; \he_a  \ =\ 
(\partial_\mu \hphi^r + A^a_\mu \,\he_a^r )\ \dd x^\mu \otimes \,
\hphi^\ast (\frac{\partial}{\partial u^r}) \ .
\label{2.3}
\end{equation}
In these formulae $\{ e_a \}$ is a basis of $\g$, $ \he_a $ is the
vector field on $X$ induced by $e_a$ and the left $G$-action, and we
have picked real coordinates $\{ x^\mu : 1\le \mu \le 2m \}$ on $M$
and $\{ u^r : 1\le r \le 2n  \}$ on $X$. Similarly, by picking complex 
 coordinates $\{ z^\alpha : 1\le \alpha \le m \}$ on $M$ and $\{ w^j :
 1\le j \le n  \}$ on $X$, one can also write down the
 anti-holomorphic part of $\dd^A \phi$ as 
\begin{equation}
\db^A \phi \ =\ \db \, \hat{\phi} \ +\ (s^\ast A^a)^{0,1} \: \he_a  \ =\ 
(\partial_{\bar{\alpha}} \hphi^j + A^a_{\bar{\alpha}}\, \he_a^j )\ \dd
\bar{z}^\alpha  \otimes \, \hphi^\ast (\frac{\partial}{\partial w^j}) \ .
\label{2.4}
\end{equation}
Having recorded these formulae we now come to the first basic fact of
the theory, namely the existence of a set of first order equations ---
the vortex equations --- whose solutions minimize  the energy
functional. This was first found in \cite{MiR} and \cite{C-G-S} for this
general non-linear sigma-model. 
\begin{thm*}[\cite{MiR, C-G-S}] 
For any connection $A \in  \AAA $ and any section $\phi \in \Gamma (E)$,
\begin{gather}
{\mathcal E} (A,\phi )\  =\ T_{[\phi ]} + \int_M \;  \| \db^A  \phi
\|^2  +  \frac{1}{2}\, \| \frac{1}{e}\, \Lambda F_A + e\ 
\mu\circ\phi  \|^2   +   \frac{2}{e^2} \,\| F_A^{0,2} \|^2     \ ,     
\label{2.5}
\end{gather}
where the term
\begin{gather}
T_{[\phi ]} \ =\  \frac{1}{2} \int_M  \frac{1}{(m-1)!} \: \phi^{\ast}
[\eta_E ] \wedge \omega_M^{m-1}\ -\   
\frac{\kappa_{ab}}{e^2 (m-2)! }\: F_A^a \wedge  F_A^b  \wedge \omega_M^{m-2}
\label{2.6}
\end{gather}
does not depend on $A$, and only on the homotopy class of $\phi $.
\label{}
\end{thm*}
\begin{cor*}[\cite{MiR, C-G-S}] 
Within each homotopy class of the sections $\phi$ we have that
${\mathcal E}(A,\phi)
\geq  T_{[\phi ]}$, and there is an equality if and only if the pair
$(A,\phi )$ in $\AAA \times \Gamma (E)$ satisfies the equations
\begin{subequations}  \label{2.7}
\begin{align}
 &  \db^A \phi \ =\  0     \label{2.7a}   \\
 &  \Lambda F_A + e^2 \ \mu\circ\phi \ = \ 0   \label{2.7b}  \\
 &  F_A^{0,2}\ =\ 0  \ \ .  \label{2.7c} 
\end{align}
\end{subequations}
These first order equations are usually called vortex equations.
\end{cor*}
Besides $\db^A \phi$, several new terms appear in (\ref{2.5}) when compared
with (\ref{2.1}); their meaning is the following. The operator $\Lambda :   
\Omega^\bullet (M) \rightarrow \Omega^{\bullet -2}(M)$ is the adjoint, with   
respect to the metric $g_M$ on $M$, of the operator $\eta \mapsto
\omega_M \wedge \eta $ 
on $\Omega^{\bullet }(M)$. By well known formulae,
\begin{gather}
\Lambda F_A \ =\ \ast (\omega_M \wedge \ast F_A )\ =\  g_M (F_A ,
\omega_M)\ ,
\label{2.8}
\end{gather}
and so $\Lambda F_A$ can be seen as a locally defined function on $M$ with
values in $\g$, just as $\mu \circ \phi$. (More properly, they should
be both regarded as global sections of $P\times_{{\rm   
Ad}_G}\g $). Next, $F_A^{0,2}$ is just the $(0,2)$-component of $F_A$ under
the usual decomposition $\Omega^2 (M) = \Omega^{2,0}\oplus \Omega^{1,1}  
\oplus \Omega^{0,2}$.
Finally $[\eta_E]$ is a cohomology class in $H^2 (E)$ that does not
depend on $A$.
Using the Cartan complex for the $G$-equivariant cohomology of $X$, $[\eta_E
]$ is just the image by the Chern-Weil homomorphism of 
the cohomology class in $H^2_G (X)$ determined by the equivariantly closed
form $\omega_X - \zeta^b \mu_b  \ \in \ \Omega_G^2 (X)$ (see for
example \cite[ch. VII]{B-G-V}).

\section{The manifold of fields}

\subsection{The manifold}

Here we continue the exposition of the previous
section by recalling some well known properties of the
infinite-dimensional manifold $\AAA \times \GME$, which is our space of
fields. Namely, we describe its natural K\"ahler structure and the action 
of the group of gauge transformations $\GG$. For more details see for 
example \cite{MiR} and the
references therein. Along the way we will also write down some explicit
expressions that will be needed in Section 4. In subsections 3.2 and 3.3 the
first basic elements of the topological gauged sigma-model are introduced,
like the BRST operator $Q$ and a few ``anticommuting fields''. These are all 
described in terms of the $\GG$-equivariant cohomology of $\AAA \times
\GME$.

$ \ $

To start the study of the manifold $\AAA \times \GME$ we first look at
its tangent space. Recall that, given a connection $A \in \AAA$, the
tangent space $T_A \AAA$ can be identified with $\Omega^1 (M; \g_P)$ --- the
space of 1-forms on $M$ with values in the bundle $\g_P := P \times_{{\rm
Ad}_G} \g $. Likewise, given a section $\phi \in \GME$, the tangent space
$T_\phi \GME$ can be identified with the space of sections of  $\phi^\ast
\ver \rightarrow M$. Here $\ver \rightarrow E$ is the sub-bundle of $TE
\rightarrow E$ defined by the kernel of $\dd \pi_E : TE \rightarrow TM$,
and $\phi^\ast \ver $ is the pull-back bundle. Thus
\[
T_{(A, \phi )} (\AAA \times \GME ) \ \simeq \ \Omega^1 (M; \g_P ) \; \oplus
\; \Gamma (\phi^\ast \ver ) \ .
\]
Both summands on the right hand side have a natural metric and complex
structure, induced by the ones on $M$ and $X$, respectivelly. Hence the
manifold $\AAA \times \GME$ has a natural metric and complex structure.
Moreover, it can be shown that this complex structure is integrable,
compatible with the metric, and that the K\"ahler form is closed. So $\AAA
\times \GME$ is a K\"ahler manifold.

More explicitly, suppose that we are given tangent vectors
\begin{align}
\tau \ &= \ \tau^{a}_{\mu} \ \dd x^\mu \otimes e_a   \quad \in \ \Omega^1
(M; \g_P )       \label{3.1}   \\
V \  &= \ V^r \ \hphi^\ast (\frac{\partial}{\partial u^r })  \quad \in \
\Gamma (\phi^\ast \ver ) \ ,   \nonumber   \label{3.2}
\end{align}
which are here written down in terms of their local representatives with
respect to trivializations of $\g_P$ and  $E$ induced by a local
trivialization of $P$. Then the K\"ahler metric on $\AAA \times \GME$ is
given by
 \[
g_{\AAA \times \GME} (\tau_1 + V_1 ,  \tau_2 + V_2 ) \ = \  \int_M
\, (\tau_1)^a_\mu \, (\tau_2)^b_\nu \, (g_M)^{\mu \nu} \,
\kappa_{ab} \
+\ e^2 \: V^r_1 \: V^s_2 \: (g_X)_{rs} \ .
\]

We now turn to the action of gauge transformations on the fields in
$\AAA \times \GME$. Recall that the group $\GG$ of gauge
transformations is the group of $G$-equivariant
automorphisms of the bundle $P \rightarrow M$ which descend to the
identity map on the base $M$. Equivalently, $\GG$ is the group of
sections of the associated bundle $P \times_{{\rm Ad}_G } G$. Using
the local trivializations (\ref{2.2}), each of these sections $g \in \GG$
is locally represented by a map from $\UU \subset M$ to $G$. The Lie
algebra of $\GG$ is the space $\Gg = \Omega^0 (M; \g_P)$ of sections
of the bundle $\g_P \rightarrow M$; each of these sections can be
locally represented by a map from $\UU$ to $\g$.

The group $\GG$ has a natural right action on the manifold of fields
$\AAA \times \GME$. For any $g \in \GG$ this action is determined by
the formulae
\begin{align}
A \cdot g  \ &=\ {\rm Ad}_g \circ A \ -\ \pi_P^\ast \, (g^{-1} \, \dd g ) \
;      \label{2.9} 
\\
\phi (x) \ &= \ [p,q] \qquad \Longrightarrow \qquad  (\phi \cdot g ) (x) \
=\ [p , \; \rho_{g_p^{-1}} (q) ] \ .  \nonumber
\end{align}
In the last formula, $\rho$ is the left $G$-action on $X$ and
$g_p$ is the only element of $G$ such that $g \circ \pi_P (p) \ =\ [p,
\; g_p] $ in $P \times_{{\rm Ad}_G} G$. For our purposes the most
relevant facts about these gauge transformations is that they preserve
the K\"ahler metric on $\AAA \times \GME$, as well as the energy
functional ${\mathcal E} (A , \phi)$ and the vortex equations.
In particular, if $(A, \phi)$ is a solution of the vortex equations, 
then so is $(A\cdot g , \phi \cdot g)$ for any $g \in \GG$.

The right action of $\GG$ on the manifold $\AAA \times \GME$ induces
linear maps from the Lie algebra $\Gg$ to the tangent spaces $T_{(A,
  \phi)} (\AAA \times \GME)$. These maps correspond to the
infinitesimal gauge transformations and are explicitly given by
\begin{align}
C_{(A, \phi)} : \ \Omega^0 (M; \g_P) &\longrightarrow \ \Omega^1(M;
\g_P) \oplus \Gamma (\phi^\ast \ver) 
\label{3.3} \\
\varepsilon = \varepsilon^a e_a \  &\longmapsto \ (\; D_A \,
\varepsilon , \; - \varepsilon^a \; \hphi^\ast (\he_a )\; ) \ .
\nonumber
\end{align}
Here $D_A$ is the covariant derivative on $\g_P \rightarrow M$ induced by
the connection $A$ on $P$, and, as explained before, $\he_a$ is the vector
field on $X$ induced by $e_a \in \g$ and the left action of $G$ on
$X$. Using the inner products on $\Gg$ and on the tangent space to $\AAA \times
\GME$, one can also consider the adjoint linear maps
$C_{(A,\phi)}^\dag = C_A^\dag \oplus C_\phi^\dag$. A standard
calculation shows that these are given by 
\begin{align}
C_A^\dag  : \ \Omega^1 (M; \g_P) &\longrightarrow \ \Omega^0 (M; \g_P)
\label{3.5}   \\
\tau \ &\longmapsto \  -(g_M)^{\mu \nu} (\; \partial_\nu \tau_\mu  -
\Gamma_{\nu \mu}^{\lambda} \tau_{\lambda} +  [A_\nu , \tau_\mu]  \; )
\nonumber 
\end{align}
and
\begin{align}
C_\phi^\dag : \ \Gamma (\phi^\ast \ver )  &\longrightarrow \ \Omega^0(M; \g_P)
\label{3.6}      \\
 V \ &\longmapsto \  -\, e^2 \, \kappa^{ab} \, (g_X)_{rs} \, (\he_b)^r
 \, V^s \, e_a  \ ,   \nonumber   
\end{align}
where the $\Gamma_{\nu \mu}^{\lambda}$'s  are the Christoffel symbols for 
the Levi-Civita connection on $M$.

Finally, there is a moment map for the right action of  $\GG$ on the
K\"ahler manifold $\AAA \times \GME$. This is a map $ \mu_{\AAA \times
\GME} : \AAA \times \GME \rightarrow \Gg^\ast$, and can be shown to be
\[ 
\mu_{\AAA \times \GME} (A, \phi ) \ = \ -\, \Lambda F_A \: -\: 
e^2 \: \mu \circ \phi \ ,
\]
where we are using the inner product on $\Gg$ to identify this space with
a subspace of $\Gg^\ast$. Notice that the second vortex equation
(\ref{2.7b}) is exactly the vanishing condition for this moment map.

\subsection{Basic differential forms}

Here we will informally define some basic ``coordinate'' functions and
differential forms on the infinite-dimensional manifold $\AAA \times
\GME$, which is our space of fields. As we will see, the definition of these 
forms depends on a choice of a local trivialization
of the bundle $P \rightarrow M$ and of a point $x$ in the domain of
this trivialization. This just means that the forms can be tensored
with sections of other appropriate bundles in order to define global
sections of a bigger bundle over the space of fields.

More concretely, one proceeds as follows.
Let $s$ be a local trivialization of the bundle $P$ over a domain $\UU$ in
$M$. Then given a connection $A \in \AAA$ and a section $\phi \in \GME$,
this trivialization allows us to pick local representatives $s^\ast A =
A^a_\mu \; \dd x^\mu  \: e_a$ and $( \ldots , \, \hphi^r , \ldots )$ for
$A$ and $\phi$; just as in (\ref{2.2}) and (\ref{2.3}). Now, keeping fixed the
trivialization $s$, the point $x\in \UU$, and the indices $a$, $\mu$
and $r$,
the maps $A \mapsto A^a_\mu (x)$ and $\phi \mapsto \hphi^r (x)$ are
actually smooth functions on $\AAA$ and on an open set of $\GME$,
respectivelly. (This is the open set of sections $\phi$ such that the
representative $\hphi (x)$ has values in the domain of the chart $\{ u^r
\}$ of $X$.) Using the exterior derivative $\tilde{\dd}$ on the manifolds
$\AAA$ and $\GME$, we can thus define the 1-forms
\begin{align}
\psi^a_\mu (x) \ &= \ \tilde{\dd} \;  [A^a_\mu (x) ]  \qquad \in \
\Omega^1 (\AAA)     \label{3.7}    \\
\chi^r (x) \ &= \ \tilde{\dd} \; [\hphi^r (x) ]  \qquad \;  \in  \ \Omega^1
({\rm open\ set\ of \ \GME }) \ .   \label{3.8}
\end{align}
It follows from the definition that, acting on the tangent vectors $\tau
\in T_A \AAA$ and  $V \in T_\phi \GME$ of (\ref{3.1}), these forms give
 \begin{align}
\psi^a_\mu (x) \; [\tau ] \ &= \  \tau^a_\mu (x)  \ ,   \label{3.9}   \\
\chi^r (x) \; [V] \  &= \ V^r (x) \ .  \nonumber %\label{3.10}   
\end{align}
These trivialization-dependent forms can be combined to define the fields
\begin{align}
\psi \ &= \  \psi^a_\mu \wedge \dd x^\mu \otimes e_a  \qquad  {\rm
  and} \label{3.11}       \\
\chi \ &= \ \chi^r \;  \hphi^\ast ( \frac{\partial}{\partial u^r }) \
  ,  \label{3.12}
\end{align}
which are global sections of the bundles
\begin{align*}
\Lambda^2 ( \AAA \times  M ) \: \otimes \: \g_P \ &\longrightarrow \  
\AAA \times M  \qquad \qquad {\rm and}    %%\label{3.13} 
   \\
 T^\ast \GME \: \otimes \: \Upsilon^\ast \ver \ &\longrightarrow \ \GME \times
M \ ,    %% \label{3.14}
\end{align*}
respectively. Here $\Upsilon^\ast \ver $ is the pull-back of $\ver
\rightarrow E$ by the natural evaluation map
\begin{equation}
\Upsilon : \GME \times M \longrightarrow E \ , \qquad  (\phi , x ) \mapsto \phi
(x)  \ .     
\label{3.15}
\end{equation}
Having in mind the expressions (\ref{3.9}), it is also clear that the
operators $C_A^\dag$ and $C_\phi^\dag$ of (\ref{3.5}) and (\ref{3.6}) can be
written as
\begin{align}
C_A^\dag  \ &= \   -\, (g_M)^{\nu \mu} \,  (D_A \psi )_{\nu \mu }^a \:
e_a  \qquad 
\qquad {\rm and}    \label{3.16} \\
C_\phi^\dag \ &= \  - \, e^2 \, \kappa^{ab}\,  (g_X)_{rs} \, (\he_b)^r
\,  \chi^s \:  e_a  \
,   \nonumber %\label{3.17}   
\end{align}   
where
\begin{gather}
(D_A \psi )_{\nu \mu }^a \ = \  \partial_\nu \psi^a_\mu  -  \Gamma_{\nu
\mu}^{\lambda} \, \psi^a_\lambda  + [A_\nu , \psi_\mu ]^a  \ .   \label{3.18}
\end{gather}

\subsection{The $\GG$-equivariant complex}

Using the differential forms defined above, let us now look into the   
$\GG$-equivariant complex of the manifold $\AAA \times \GME$. This will
lead to the definition of the BRST operator $Q$.

For the sake of clarity we first recall a finite-dimensional
example, for instance the $G$-equivariant complex of $X$ in the Cartan
model \cite{B-G-V, Gu-S}. This complex is defined as the space
\begin{equation*}
\Omega_G  (X) \ := \  \left(  S^\bullet (\g^\ast ) \otimes  \Omega^\bullet
(X)  \right)^G
\end{equation*}
of $G$-invariant elements in the tensor product of the symmetric algebra $
S^\bullet (\g^\ast )$ with the de Rham algebra $\Omega^\bullet (X)$. The  
differential operator acting on this space is defined to be $\dd_{\mathcal
C} = \dd \;- \; e^a \otimes \iota_{-\he_a}$, where $\dd$ is the exterior
derivative on $X$, $\{ e^a  \}$ is the basis of $\g^\ast$ dual to the
basis $\{ e_a \}$ of $\g$, and $- \he_a $ are the vector fields on $X$
induced by $e_a$ and the right $G$-action. Notice also that, for any
$\alpha \in \Omega^\bullet (X) $ and $v \in \g$,
\begin{gather}
( e^a \otimes \iota_{-\he_a} \ \alpha  ) \ [v] \ = \ \iota_{- v^a \he_a}
\alpha \ = \ \iota_{C(v)} \alpha \ ,
 \label{3.19} 
\end{gather}
where $C : \g \rightarrow \Gamma (TX)$ is the linear map induced by the
right $G$-action on $X$.

In the case of the infinite-dimensional manifold $\AAA \times \GME$ with
right $\GG$-action, this picture becomes the following. The
$\GG$-equivariant complex $\Omega^\bullet_{\GG} (\AAA \times \GME )$ is
the space of $\GG$-invariant sections of the bundle
\begin{equation}
S^\bullet (\Gg^\ast ) \; \otimes \; \Lambda^\bullet (\AAA \times \GME ) \
\longrightarrow \ \AAA \times \GME \ ,
 \label{3.20}
\end{equation}
where the first factor in the tensor product is the trivial bundle over
$\AAA \times \GME$ with fibre $ S^\bullet (\Gg^\ast ) $, and the second
factor is the exterior bundle of  the base. The appropriate analog of the
differential $\dd_{\mathcal C}$, which will be specified below, is then
what is usually called the BRST operator $Q$.

To define more explicitly the operator $Q$ we start by introducing the
elements $ \varphi^a (x) \; \in \Gg^\ast $, which are defined by
\begin{gather}
\varphi^a (x) \ [v] \ = \ v^a (x)  \ \qquad {\rm for\ any} \ \quad  v\ =\ v^a 
(x) \; e_a \ \in \ \Omega^0 (M; \g_P) \ .
 \label{3.21}
\end{gather}
These elements depend on the choice of local trivialization, but can be
combined to define the field
\begin{equation}
\varphi \ = \  \varphi^a (x) \; e_a \ ,
 \label{3.22}
\end{equation}
which is a global section of the bundle
\begin{equation*}
\Gg^\ast \; \otimes \; \g_P \ \longrightarrow \ M \ .
%% \label{3.23}
\end{equation*}
Now, having in mind the definition of $\dd_{\mathcal C}$, it is clear that
the analog $Q$ must act on functions on $\AAA \times \GME$ just like
$\tilde{\dd}$, the exterior derivative. It must also annihilate $\varphi^a
(x) $, just as $\dd_{\mathcal C}$ annihilates $e^a$. Thus
\begin{align}
Q\ A^a_\mu (x) \ &= \ \psi^a_\mu (x)\ ;  \qquad \quad   Q\ \varphi^a
(x) \ = \ 0 \ ;  \label{3.24}     \\    
Q\  \hphi^r (x) \ &= \ \chi^r (x)  \ .   \nonumber
\end{align}
Furthermore, using expressions (\ref{3.3}), (\ref{3.9})
and (\ref{3.21}),  
one has that for any $v \in \Omega^0 (M; \g_P )$,
\begin{align*}
\iota_{C(v)} \; \psi^a_\mu (x) \ \mid_A \  &= \  (D_A v)^a_\mu (x) \ = \
(D_A \varphi )^a_\mu (x) \ [v] \ , \\
\iota_{C(v)} \; \chi^r (x)  \ \mid_\phi \ &= \ - v^a (x) \  \he^r_a \circ
\hphi (x) \ =\ - \varphi^a (x) \;  \he^r_a \circ \hphi (x) \ [v] \ .
\end{align*}
So it follows from the definition of $\dd_{\mathcal C}$, (\ref{3.19})
and the identity 
$\tilde{\dd}^2 = 0$ that
\begin{align}
Q \ \psi^a_\mu (x) \ &= \ -(D_A \varphi )^a_\mu (x) \ ;      \label{3.25}   \\
Q\ \chi^r (x) \ &= \  \varphi^a (x) \ \he^r_a \circ \hphi (x) \ .
\nonumber 
\end{align}
Just as the Cartan operator $\dd_{\mathcal C}$, one has that $Q^2 = 0$
when acting on $\GG$-invariant sections of the bundle (\ref{3.20}). When
acting on other sections, such as the $A^a_\mu (x)$, $Q^2$ is just like an
infinitesimal gauge transformation parametrized by $\varphi$.

\section{The topological action}

The aim of this section is to write down an expression for the action of
the topological gauged sigma-model. The approach is a practical one: we
introduce the necessary fields, explain what calculations should be
performed, and spell out the final answer in (\ref{4.11}) and
(\ref{4.13}). As explained in the 
Introduction, underlying our calculations there is a more fundamental
geometrical picture, which justifies the numerous apparently arbitrary  
choices made here. For more details on this geometrical picture we refer
the reader to Appendix A and the reviews \cite{C-M-R, Bl}.

$\ $

The conventions used here are the following. The greek indices $\mu , \nu
, \rho  \ldots $ and $\alpha , \beta , \gamma \ldots $  refer to real and
complex charts, respectivelly $\{ x^\mu : \; 1\le \mu \le 2m  \}$ and $\{ 
z^\alpha : \; 1\le \alpha \le m \}$, on the manifold $M$. The latin
indices $r, s, t \ldots $ and  $i, j, k \ldots$ refer to real and complex
charts, respectivelly and $\{ u^r : \; 1\le r \le 2n \}$ and $\{ w^j : \;
1\le j \le n \}$, on the manifold $X$. The relations between the real and
complex coordinates are the usual ones
\begin{gather*}
z^\alpha \ = \ x^{2\alpha -1} + i x^{2\alpha} \qquad {\rm and}  \qquad
w^j \ =\ u^{2j-1} + i u^{2j} \ .
\end{gather*}
Just as in the real case of Section 3, the complex charts on $M$ and
$X$ induce complex coordinates and forms on $\AAA \times \GME$. These
are related to the real ones by
\begin{align}
\hphi^j_\CC \ &= \ \hphi^{2j-1} + i \hphi^{2j}\ ;   \quad  & (A_\CC
)_\alpha^a \ =\ 
\overline{(A_\CC )^a_{\bar{\alpha}} } \ = \  (A^a_{2\alpha - 1} - i
A^a_{2\alpha })/2  \ ;  \label{4.1}  \\
\chi^j_\CC \ &= \ \chi^{2j-1} + i \chi^{2j} \ ;  \quad  & (\psi_\CC
)_\alpha^a \ =\ 
\overline{(\psi_\CC )^a_{\bar{\alpha}} } \ = \  (\psi^a_{2\alpha - 1}
- i \psi^a_{2\alpha })/2  \ . \nonumber
\end{align}
In the future we will omit the subscript $\CC$ and use the type of
indices to distinguish real from complex; as for the charts, $\mu$ and
$r$ means real, $\alpha$ and $j$ means complex, etc.

Regarding the K\"ahler geometry of $M$ and $X$, we always work with the
holomorphic tangent bundles, not the complexified ones. The hermitian
metric $h$ is related to the real metric and the K\"ahler form by

\[
h \ = \ h_{\alpha \bar{\beta}} \; \dd z^\alpha  \otimes \dd \bar{z}^\beta
\ = \ g \; -\; i \; \omega \ .
\]
The hermitian (Levi-Civita) connection satisfies
\[
\nabla_{ \frac{\partial}{\partial z^\alpha}} {\frac{\partial}{\partial
z^\beta}} \ = \ \Gamma_{ \alpha \beta }^\gamma \; \frac{\partial}{\partial
z^\gamma} \ =\ h^{\gamma \bar{\delta}}\, (\partial_\alpha  h_{\beta
\bar{\delta}}) \;\frac{\partial}{\partial z^\gamma}  \ ,
\]
and the curvature components are
\[
R_{\alpha \bar{\beta} \delta \bar{\gamma}} \ = \ - \partial_\delta
\partial_{\bar{\gamma}} h_{\alpha \bar{\beta}} \ +\  h^{\varepsilon
\bar{\eta}} \, (\partial_{\delta}  h_{\alpha \bar{\eta} } ) \:
(\partial_{\bar{\gamma}}  h_{\varepsilon \bar{\beta}} ) \ .
\]
The type of indices used distinguishes whether we are working on $M$ or on
$X$.

$\ $

Having stated the conventions, we will now construct the topological
action. Firstly we need to introduce several new fields, the so-called
antighosts. These are the fields
\begin{gather*}
b_{\bar{\alpha} \bar{\beta}}^{a} (z) , \qquad c^a (z), \qquad
d^j_{\bar{\alpha}} (z),  \qquad  \lambda^a (z),
\end{gather*}
of respective ghost number $-1$, $-1$, $-1$ and $-2$, and their partners
\begin{gather*}
B_{\bar{\alpha} \bar{\beta}}^{a} (z) , \qquad C^a (z), \qquad
D^j_{\bar{\alpha}} (z),  \qquad  \eta^a (z),
\end{gather*}
of respective ghost number $0$, $0$, $0$ and $-1$. The BRST operator $Q$
acts on these fields according to the rules
\begin{align}
&Q\; b_{\bar{\alpha} \bar{\beta}}^{a} \ = \ B_{\bar{\alpha}
\bar{\beta}}^{a} \ ; \qquad \qquad  \qquad \quad \ \  
 Q\;  B_{\bar{\alpha} \bar{\beta}}^{a} \ = \ f^a_{bc} \; \varphi^b \;
b_{\bar{\alpha} \bar{\beta}}^{c}\ ;    \nonumber \\
&Q\; c^a \ = \ C^a \ ;  \qquad  \qquad \qquad \quad\ \ \ \ \ Q\; C^a \
=\ f^a_{bc} \; \varphi^b \; c^c \ ;  \nonumber
\\
&Q\; d^j_{\bar{\alpha}} \ = \ D^j_{\bar{\alpha}} \ -\ \Gamma_{ik}^j \;
\chi^k \; d^i_{\bar{\alpha}} \ ;   \nonumber  \\
&Q\; D^j_{\bar{\alpha}} \ = \ R_{i\bar{k}l\bar{m}} \; h^{j\bar{k}} \chi^l \, 
\overline{\chi^m} d^i_{\bar{\alpha}} \ - \ \Gamma_{kl}^j D^k_{\bar{\alpha}}
\chi^l \ +\  \varphi^a (\nabla \he_a)^j_k d^k_{\bar{\alpha}} \ ;  \nonumber
\\
&Q\; \lambda^a \ = \ \eta^a \ ;  \qquad  \qquad \qquad \quad\ \ \ \ \
Q\; \eta^a  \ =\ f^a_{bc} \; \varphi^b \; \lambda^c \ ;
\label{4.2}
\end{align}
where the $f^a_{bc}$ are the structure constants of the Lie algebra $\g$.
A geometric interpretation of the antighost fields and of this $Q$-action
is given in Appendix A. Here we only remark that the relations
\[
Q\; b \ =\ B \ , \qquad Q\; c \ =\ C \ , \qquad Q\; d \ =\
D - \ \cdots \qquad {\rm and} \qquad Q\; \lambda \ =\ \eta 
\] 
should be regarded as defining the fields $B$, $C$ and $D$. In particular,
it is the field $D$ that depends explicitly on the metric of the manifold
$X$, not the operator $Q$. In fact, $Q$ is a geometric operator that only
depends on $M$ and on the $G$-manifold $X$.

$\ $

As is usually the case with cohomological field theories, the action for
our model will be $Q$-exact, i.e. will be of the form
\[
I \ = \ Q\ \Psi \ ,
\]
where $\Psi$ is the so-called gauge fermion. This gauge fermion can be
split into two parts
\begin{equation}
\Psi \ =\  \Psi_{\rm localization} \ +\ \Psi_{\rm projection} \ ,
\label{4.3}
\end{equation}
which play different roles in the geometric interpretation of the
action. Moreover, in cohomological field theories there is a fairly
standard procedure to construct explicit expressions for the gauge
fermions. This is reviewed for example in \cite{C-M-R}. Going through that
procedure in the case of our gauged sigma-model, one gets at the end
\begin{align*}
\Psi_{\rm loc} \  = \ &\pm  i\, (\; d\; , \; \db^A \phi \;) \ \pm \
\frac{i}{\sqrt{2} e} \, (\; c \; , \; \Lambda F_A +  e^2 \mu \circ
\phi  \; ) \ \pm 
\ \frac{\sqrt{2}i}{e}\, (\; b \; , \; F_A^{0,2} \; ) \ +  %%  \label{4.4} 
  \\
&+ \ t\, (\: b \: ,  \: B \: ) \ +\ t\, (\: c\: , \: C
\: ) \ + \ t\, (\: d\: , \: D \: ) \ ;  \nonumber  \\
\Psi_{{\rm proj}}\ = \  &-\: i \, (\; \lambda \; , \; C_A^\dag +
C_\phi^\dag \; ) \ ;  %%  \label{4.5}
\end{align*} 
where in the first expression $t$ is an arbitrary positive parameter and
there are two possible choices of signs.
The pairings $(\cdot ,  \cdot )$ are the natural inner products on the
respective spaces. Explicitly, using complex coordinates,
\begin{align}
(\; d\; , \; \db^A \phi \;) \ &= \ 2 \,\Re {\mathfrak e}  \int_M
\overline{d^j_{\bar{\alpha}}} \, (\db^A \phi )^k_{\bar{\beta}} \,  h^{\alpha 
\bar{\beta}}\,  h_{k\bar{j}}  \ ,         \label{4.6}     \\
(\; c \; , \; \Lambda F_A +  e^2 \, \mu \circ \phi  \; ) \ &= \ \int_M   c^a
\; (  \Lambda F_A^b +  e^2  \kappa^{bc} \; \mu_c \circ \phi   ) \
\kappa_{ab}   \ ,       \label{4.7}         \\
(\; b \; , \; F_A^{0,2} \; ) \ &= \  2 \,\Re {\mathfrak e}  \int_M
b_{\bar{\alpha} \bar{\beta}}^a \:  \overline{(F_A)^c_{\bar{\gamma}
\bar{\delta}}} \; h^{\gamma \bar{\alpha}}\, h^{\delta \bar{\beta}}\, 
\kappa_{ac}  \ ,             \label{4.8}
\end{align}
and the expressions for $(c , C)$, $(d, D)$ and $(b, B)$ are analogous.
Rewritting the operators $C^\dag$ of (\ref{3.16}) in complex coordinates, we
also have that
\begin{gather}
(\; \lambda \; , \; C_A^\dag + C_\phi^\dag \: ) \ = \ \int_M \lambda^a
 \, \kappa_{ab} \ \Re {\mathfrak e}\left[  -\, 4 h^{\alpha
 \bar{\beta}}\, (D_A \psi )^b_{\alpha \bar{\beta}} \; - \; e^2\,
 \kappa^{bc}\,  h_{j\bar{k}} \, (\he_c)^j \,    
\overline{\chi^k}    \right]  \ ,     
\label{4.9}
\end{gather}
where
\begin{gather}
(D_A \psi )^a_{\alpha \bar{\beta}} \ = \  \partial_{\alpha} \,
 \overline{\psi^a_\beta}  
 + f_{bc}^a \: A^b_\alpha \: \overline{\psi^c_\beta} \ .   \label{4.10}
\end{gather}

The final step is to go from the gauge fermion to the Lagrangian, and this
is just a computational matter. One acts with the operator $Q$ on $\Psi$ and
integrates out the auxiliary fields $B$, $C$ and $D$. A few important
intermediary stages in this calculation are the following. The section
$\Lambda F_A$ can be written in real and complex coordinates as
\[
\Lambda F_A^a \ = \ \frac{1}{2}\, g^{\mu \sigma}\, g^{\nu \lambda}\,
\omega_{\mu \nu}\,  (F_A^a)_{\sigma \lambda} \ = \ 2\, \Im {\mathfrak
  m} [\: h^{\alpha
\bar{\beta}}\,  (F_A^a)_{\alpha \bar{\beta}} \:  ]  \ ,
\]
where
\[
F_A^a \ = \ \frac{1}{2} \, (F_A^a)_{\mu \nu} \; \dd x^\mu  \wedge \dd x^\nu \
= \  \Re {\mathfrak e}\: [\: (F_A^a)_{\alpha \beta} \; \dd z^\alpha  \wedge
\dd z^\beta  \; + \;   (F_A^a)_{\alpha \bar{\beta}} \; \dd z^\alpha
\wedge \dd \bar{z}^\beta \: ] \ .
\]
So
\[
Q\; \Lambda F_A^a \ = \  4 \, \Im {\mathfrak m} \:[\: h^{\alpha
    \bar{\beta}}\, (D_A \psi )^a_{\alpha \bar{\beta}} \: ] \ .
\]
Using the definition of  moment map,
\[
Q \; (\mu_c \circ \phi ) \ = \ \Im {\mathfrak m} \: [\: h_{j\bar{k}}\,
  (\he_c)^j \, \overline{\chi^k} \: ]  \ .
\]
Furthermore, using the holomorphy of the vector fields $\he_a$ and the
properties of the hermitian connection $\nabla$ on $X$, one gets that
\begin{equation*}
Q \ [\: (\db^A \phi )^j_{\bar{\alpha}} \, h_{j\bar{k}}\: ] \ = \ \left[\: (   
(\phi^\ast \nabla^A)^{0,1} \chi )^j_{\bar{\alpha}} + \psi^a_{\bar{\alpha}}\,
  (\he_a)^j \:  
\right] \: h_{j\bar{k}} \ +\  (\db^A \phi )^j_{\bar{\alpha}}\,
h_{j\bar{l}}\, 
\overline{\Gamma_{km}^l \, \chi^m }  \ . 
\end{equation*}
In this last expression
\begin{equation}
[\: (\phi^\ast \nabla^A)^{0,1} \chi\: ]^j_{\bar{\alpha}} \ = \
\partial_{\bar{\alpha}} 
\chi^j \; + \; A^a_{\bar{\alpha}}\, \chi^k \, (\nabla \he_a )^j_k \;
+ \; \Gamma_{kl}^j \, 
(\partial_{\bar{\alpha}} \hphi^k ) \,  \chi^l
\label{4.101}
\end{equation}
is the anti-holomorphic part of the connection $\phi^\ast \nabla^A$ on
the bundle 
$\phi^\ast \ver \rightarrow M$ induced by $A$ and the hermitian connection
on $X$. All throughout the calculation one should also bear in mind that
functions on $X$ such as $h_{j \bar{k}}$ or $(\he_a)^j$ depend implicitly
on $\phi$, because they are to be evaluated at the point $\hphi (x)$, with
$x \in M$. This implies for example that
\[
Q\ h_{j \bar{k}} \ =\ (\partial_l  h_{j \bar{k}} ) \, \chi^l  \ + \
(\partial_{\bar{l}} h_{j \bar{k}} )\, \overline{\chi^l } \ .
\] 

At the end of the calculation, what we get for the localization part of
the action is
\begin{align}
I_{{\rm loc}} \ = \ &\frac{1}{4t}\, \left[ \: \|  \db^A \phi  \|_M^2 \: +\:
\frac{1}{2 e^2}\, \| \Lambda F_A +  e^2 \mu \circ \phi \|_M^2 \:   + \:
\frac{2}{e^2}\,  \|  F_A^{0,2}  \|_M^2  \:  \right] \: \mp    \label{4.11}    \\
&\mp  \ i\  ( \; d \; , \;  (\phi^\ast \nabla^A)^{0,1} \chi   +
\psi^a_{\bar{\alpha}} \; \dd \bar{z}^\alpha  \otimes \he_a \; ) \ \mp \
\frac{\sqrt{2} i}{e} \ (\; b \; , \; (D_A \psi)^{0,2} \; )  \ \mp
\nonumber \\
&\mp \ \frac{i}{\sqrt{2}e} \ \int_M    c^a  \  \Im {\mathfrak m}
\left[ \, 
4 \, \kappa_{ab} \, h^{\alpha \bar{\beta}}\,  (D_A \psi )^b_{\alpha
  \bar{\beta}}  \;
- \; e^2  h_{j\bar{k}} \,(\he_a)^j \,  \overline{\chi^k} \:   \right] \ -
\nonumber \\
&- \ t \: (\; b \; , \; [\varphi , b ] \; ) \  - \ t
\: (\; c \; , \; [\varphi , c ] \; ) \ +   \nonumber    \\
&+ \ 2\, t \ \int_M  R_{i\bar{j}k\bar{l}}\, d^i_{\bar{\alpha}}\, 
\overline{d^j_{\bar{\beta}}} \, \chi^k \, \overline{\chi^l} \,  h^{\beta
\bar{\alpha}}\; +\; \Re {\mathfrak e} \left[\: \varphi^a \, (\nabla \he_
  a)^k_i \, d^i_{\bar{\alpha}}\, \overline{d^j_{\bar{\beta}}} \,
  h_{k\bar{j}}\,  h^{\beta \bar{\alpha}} \:   \right] \ ,  \nonumber
\end{align}
where $(\cdot , \cdot)$ are the natural inner  products of
(\ref{4.6})-(\ref{4.8}), and   
the norms $\|  \cdot \|_M $ come from these inner products. In the
expression above we have already integrated out the fields $B$, $C$ and 
$D$. 
%%\begin{align}
%%B \ &= \  \mp  \frac{1}{\sqrt{2} t}\,  F_A^{0,2} \ ; \qquad \quad
%%&D\ =\ \mp 
%%\frac{1}{2t}\,  \db^A \phi \ ;      \label{4.12}      \\
%%C \ &= \ \mp \frac{1}{2 \sqrt{2} t} \: ( \Lambda F_A +  e^2 \, \mu
%%\circ \phi ) 
%%\ .  \nonumber
%%\end{align}
Observe that, up to a factor, the bosonic part of $I_{{\rm loc}}$
 is equal to the classical energy
minus the energy $T_{[\phi]}$ (see expression (\ref{2.5})). This
quantity is minimized exactly by the vortex solutions.

The calculations for the projection part of the action give
\begin{align}
I_{{\rm proj}}\ = \ &- \, i\:  (\; \eta \; , \; C_A^\dag +
C_\phi^\dag \; ) \ +       \label{4.13}     \\
&+  i \int_M  \lambda^a \ \Re {\mathfrak e} \Big[ \: 4\, 
\kappa_{ab} \, h^{\alpha \bar{\beta}} \: (\, f^a_{bc}\, \psi^b_\alpha \, 
\overline{\psi_\beta^c }\:  - \: (D_A D_A \varphi )^b_{\alpha
  \bar{\beta}}\, 
)   \nonumber      \\   
&+ \ e^2 \, h_{j\bar{k}} \, (\he_a)^j \,  \overline{(\he_c)^k} \,
\varphi^c \ +\ e^2 \, h_{j\bar{k}}\, (\nabla \he_a )^j_l \, \chi^l \,
\overline{\chi^k} \ \Big] 
\ , \nonumber 
\end{align}
where the first term is similar to (\ref{4.9}) and
\[
(D_A D_A \varphi )^b_{\alpha \bar{\beta}} \ = \ \partial_{\bar{\beta}}
(D_A \varphi )^b_\alpha \ +\ f_{cd}^b \, A^c_\alpha\,  \overline{(D_A \varphi
)_\beta^b } \ .
\]

\section{Observables I --- definition}

\subsection{The homomorphism $\OO$}

After having defined the field content and Lagrangian of our theory,
the next natural step is to find an interesting set of observables
whose correlation functions we would like to compute. The purpose of
this section is then to define one such a set. Observables are by
definition $Q$-closed elements of $\Omega_\GG^\bullet (\AAA \times
\GME)$ --- the equivariant complex of the space of fields. In this
section, roughly speaking, we will define one observable for each
given element of $\Omega^\bullet_G (X)$ --- the equivariant complex of
$X$. The construction presented here just combines into a single
formalism the constructions of observables given in \cite{W1} for
topological Yang-Mills and in \cite{W2} for sigma-models coupled to
gauge fields.

$\ $

Consider the trivial extension of the $\GG$-action on $\AAA \times \GME$
to the product manifold $\AAA \times \GME \times M$, and denote by
$\Omega^\bullet_\GG (\AAA \times \GME \times M)$ the associated
$\GG$-equivariant complex. Recall that, as a vector space, this complex is
just the space of $\GG$-invariant sections of the bundle
\begin{equation}
S^\bullet (\Gg^\ast )\; \otimes \; \Lambda^\bullet (\AAA \times \GME \times
M) \ \longrightarrow \ \AAA \times \GME \times M \ .
\label{5.1}
\end{equation}
The first step towards defining our set of observables will be to  
construct a homomorphism of complexes
\begin{equation}
\OO : \ \Omega^\bullet_G (X) \ \longrightarrow \ \Omega^\bullet_\GG (\AAA
\times \GME \times M ) \ .
\label{5.2}
\end{equation}
This construction involves the sections $\varphi$, $\psi$ and $\chi$
defined in (\ref{3.22}), (\ref{3.11})  and (\ref{3.12}),
respectivelly, as well as the new sections
\begin{align}
F : \ \AAA \times M  \ &\longrightarrow \  \Lambda^2 (M) \; \otimes \; \g_P
\label{5.23}   \\
(A,  x)  \ &\longmapsto \  (F_A)^a (x) \ e_a   \nonumber
\end{align}
and
\begin{align}
{\mathbb D} : \ \AAA \times \GME \times M  \ &\longrightarrow \  \Lambda^1
(M) \; \otimes \; \Upsilon^\ast \ver    \label{5.24}  \\
(A, \phi, x)  \ &\longmapsto \ (\dd^A \hphi)^j (x) \  \hphi^\ast
(\frac{\partial}{\partial w^j}) \ .    \nonumber
\end{align}
In these formulae $F_A$ is the curvature of the connection $A$, and $\dd^A
\phi$ is the covariant derivative of expression (\ref{2.3}). Notice also that
both $\chi$ and ${\mathbb D}$ can be regarded as sections of the ``bigger''
bundle
\begin{equation}
\Lambda^\bullet (\AAA \times \GME \times M) \; \otimes \; \Upsilon^\ast \ver
\ \longrightarrow  \ \AAA \times \GME \times M \ ,
\label{5.21}
\end{equation}
while $\varphi$, $\psi$ and $F$ can be regarded as sections of
\begin{equation}
S^\bullet (\Gg^\ast )\; \otimes \; \Lambda^\bullet (\AAA \times \GME \times
M) \; \otimes \; \g_P \ \longrightarrow \ \AAA \times \GME \times M \ .
\label{5.22}
\end{equation}

The  homomorphism $\OO$ can now be defined as follows. Let $\alpha$ be any
element of the complex $\Omega^\bullet_G (X) = [S^\bullet (\g^\ast) 
\otimes \Omega^\bullet (X)]^G$. It can be locally written as
\begin{equation}
\alpha \ = \ \frac{1}{k!\, l!} \; \alpha_{a_1 \cdots a_k r_1 \cdots r_l }
(u) \ \zeta^{a_1} \cdots \zeta^{a_k} \; \dd u^{r_1} \wedge \cdots \wedge 
\dd u^{r_l} \ ,
\label{5.3}
\end{equation}
where $u \in X$ and the coefficients $\alpha_{a_1 \cdots a_k r_1 \cdots
r_l }$ are symmetric on the $a_j$'s and anti-symmetric on the $r_j$'s.
Then the section $\OO_\alpha \; \in \;  \Omega^\bullet_\GG (\AAA \times
\GME \times M)$ is defined by the local formula
\begin{equation}
\OO_\alpha (A, \phi , x ) \ = \ \frac{1}{k!\, l!} \; (\alpha_{a_1 \cdots a_k
r_1 \cdots r_l } \circ \hphi ) \ \left[\prod_{j=1}^k    (\varphi + \psi + F_A
)^{a_j} \right] \ \left[ \prod_{i=1}^l  (\chi + \dd^A \hphi )^{r_i} \right] \ ,
\label{5.4}
\end{equation}
where, on the right hand side, we have omitted the dependence on $x \in   
M$.

$\ $

It is not obvious a priori that the homomorphism $\OO$ is globally well
defined. This is because the local components $(\varphi + \psi + F )^{a}$,
$(\chi + \dd^A \hphi )^{r}$ and $\alpha_{a_1 \cdots a_k r_1 \cdots r_l }
\circ \hphi $ depend on the choice of trivialization of $P$, which
determines the trivializations of  $\g_P$ and $E$. Furthermore, one should
also check the invariance of $\OO_\alpha$ under the $\GG$-action on the
bundle (\ref{5.1}). We will now sketch how all this is done.

Consider a gauge transformation $g \in \GG$. It can be locally represented
by maps $\hat{g} : \UU \rightarrow G$, where $\UU$ is a domain in $M$. One
needs to compute the transformation rules of the components $(\varphi +
\psi + F )^{a}$ and  $(\chi + \dd^A \hphi )^{r}$ under the action of $g$.
Notice as well that, since a local gauge transformation is equivalent to a
local change of trivialization of $P$ (determined  by the transition
function $\hat{g}$), these rules coincide with the transformation rules of
the various components under change of trivialization of $P$.

Let us start with the fields $\varphi$, $\psi$ and $F$, which are sections
of the bundle (\ref{5.22}). The left $\GG$-action on this bundle is induced by
the coadjoint action on $\Gg^\ast$, the pull-back action on
$\Omega^\bullet (\AAA \times  \GME \times M )$, and the usual action on
$\g_P$. Using the respective definitions one can compute that, under the  
action of  $g\in \GG$, the components of the fields transform as
\begin{align*}
\varphi^a (x) \ &\rightarrow \   ({\rm Ad}_{\hat{g}(x)^{-1}} )^a_b \
\varphi^b (x) \ ; \qquad \quad  F^a (x) \ \rightarrow \  ({\rm
Ad}_{\hat{g}(x)^{-1}} )^a_b \ F^b (x) \ ;  \\
\psi^a_\mu (x) \  &\rightarrow \ ({\rm Ad}_{\hat{g}(x)^{-1}} )^a_b \
\psi^b_\mu  (x) \ .
\end{align*}
On the other hand, the local sections $e_a (x)$ of $\g_P$ transform as 
\[
e_a (x) \ \rightarrow \ ({\rm Ad}_{\hat{g}(x)} )_a^b \  e_b (x) \ .
\]
This makes apparent the following two facts. Firstly, regarding $\hat{g}$ 
as a transition function, the sections $\varphi$, $\psi$ and $F$ defined
by (\ref{3.22}),  (\ref{3.11}) and  (\ref{5.23}) are well defined, i.e. are
trivialization independent. Secondly, regarding $\hat{g}$ as a local gauge
transformation, the sections $\varphi$, $\psi$ and $F$ are
$\GG$-invariant.

The remaining fields $\chi$ and ${\mathbb D}$ are sections of the bundle 
(\ref{5.21}). The left $\GG$-action on this bundle is induced by the pull-back
action on $\Omega^\bullet (\AAA \times \GME \times M)$ and the
push-forward action on $\Upsilon^\ast \ver$. Using the respective
definitions one can compute that, under the action of  $g \in \GG$, the
components of these fields tranform as
\begin{align*}
\chi^{\tilde{r}} (x) \ &\rightarrow \ ( \dd  \rho_{\hat{g} (x)^{-1}}
)^{\tilde{r}}_s  \circ \hphi (x) \ \; \chi^s (x )  \ ;  \\
{\mathbb D}^{\tilde{r}} (x) \ &\rightarrow \  ( \dd  \rho_{\hat{g}
(x)^{-1}}  )^{\tilde{r}}_s  \circ \hphi (x) \ \; {\mathbb D}^s (x )    \ ;   
\end{align*}
where the tilde over the index $r$ allows for a possible change of chart
on the target $X$. On the other hand the local sections of $\Upsilon^\ast
\ver$ transform as
\[
(g^{-1} \cdot \phi )^\ast_x   \; (\frac{\partial}{\partial \tilde{u}^r}) \
\rightarrow \ ( \dd  \rho_{\hat{g} (x) }  )^s_{\tilde{r}}  \circ ( g^{-1}
\cdot \phi ) (x) \ \; \phi^\ast (\frac{\partial}{\partial u^s})    \ .
\]
As before, this makes apparent that $\chi$ and ${\mathbb D}$ are globally
well defined as sections of the bundle (\ref{5.21}), and that, moreover, they
are $\GG$-invariant.
Finally, substituting all the transformation rules into expression
(\ref{5.4}), which defines $\OO_\alpha$, one can compute that
\[
(g \cdot \OO_\alpha )_{(A, \phi , x )} \ = \ (\OO_{ \hat{g}(x) \cdot
\alpha })_{(A , \phi , x)} \ = \ (\OO_{\alpha})_{(A , \phi , x)} \ .
\]
Here the notation $\hat{g}(x) \cdot \alpha $ refers to the natural
$G$-action on $S^\bullet (\g^\ast) \otimes \Omega^\bullet (X)$, and in  
the last equality we have used that, by assumption, $\alpha$ is
$G$-invariant. As before, this shows at the same time that $\OO_\alpha $
is well defined and $\GG$-invariant.

$\ $

Up to now we have only established that the map $\OO$ of (\ref{5.2}) is well
defined. Since the claim is that $\OO$ is a homomorphism of  complexes, we
must also show that it intertwines the differential operators.

By definition, the differential operator on the $\GG$-equivariant complex
of $\AAA \times \GME$ is the operator $Q$, presented in Section 3. Thus
for the trivial extension of the $\GG$-action to $\AAA \times \GME \times
M$, the differential operator on the complex $\Omega^\bullet_\GG (\AAA
\times \GME \times M )$ is $\dd_M + Q$, where $\dd_M$ denotes the exterior
derivative on $M$ regarded as acting on forms over $\AAA \times \GME
\times M$. Calling $\dd_{\mathcal C}$ the usual equivariant differential  
on $\Omega^\bullet_G (X)$, our aim is to show that
\begin{equation*}
\OO_{\dd_{\mathcal C} \alpha} \  = \ (\dd_M + Q ) \; \OO_\alpha
%%\label{5.5}
\end{equation*}
for all $\alpha$ in $\Omega^\bullet_G (X)$. This implies in particular
that $\OO$ induces a homomorphism of cohomology groups $H^\bullet_G (X)
\rightarrow  H^\bullet_\GG (\AAA \times \GME \times M)$.  
Having in mind the definition (\ref{5.4}) of $\OO$, the first step is to see
how $\dd_M + Q$ acts on the fields $\varphi$, $\psi$, $F$, $\chi$ and
${\mathbb D}$. This computation requires the formulae of (\ref{3.24})
and (\ref{3.25}). After some
algebra and several cancelations one gets
\begin{align*}
(\dd_M + Q) \;  (\varphi + \psi + F_A )^a \ &= \ - \, \dd x^\mu \ [A_\mu ,    
\varphi + \psi + F_A ]^a  \ ; \\
(\dd_M + Q)  \; (\chi + \dd^A \hphi )^r \ &= \ (\varphi + \psi + F_A)^a 
(\he_a)^r \ - \ A^a \; \partial_s (\he_a)^r \; (\chi + \dd^A \hphi )^s \ .
\end{align*}
This computation also uses the identity
\[
A^a \, A^c \,(\partial_s \he_a^r )\, \he_c^s \ = \ \frac{1}{2}\, [A, A]^a \;
\he_a^r \ ,
\]
which follows from the usual formula $[\he_a , \he_c ]^r  = - f_{ac}^b
\he_b $. Applying these formulae to the definition (\ref{5.4}) of
$\OO_\alpha$, a rearrangement of terms shows that
\[
(\dd_M + Q ) \ \OO_\alpha \ = \ \OO_{\dd_{\mathcal C} \alpha } \ +\ A^a \;
\OO_{e_a \cdot \alpha} \ ,
\]
where $e_a \cdot \alpha $ refers to the representation of $\g$ on
$S^\bullet (\g^\ast) \times \Omega^\bullet (X)$ induced by the right
$G$-action on this space. Since by assumption $\alpha \in
\Omega^\bullet_G (X)$ is $G$-invariant, we have that $e_a \cdot \alpha =
0$, and so the result follows.

\subsection{Natural observables}

Observables of our topological field theory are, by definition, $Q$-closed
elements of  $\Omega^\bullet_\GG (\AAA \times \GME )$. Thus an observable
determines a cohomology class in $H^\bullet_\GG (\AAA \times \GME )$.   
Making use of the homomorphism $\OO$ defined above, it is now
straightforward to construct a large set of observables for our theory.
This construction goes just as in references \cite{W1, W2}.
  
$\ $

Let $\alpha \in \Omega^{\bullet}_G (X)$ be any equivariantly closed form,
and consider its image $\OO_\alpha \; \in  \; \Omega^\bullet_\GG (\AAA
\times \GME \times M )$, which is $(\dd_M + Q)$-closed. Decomposing
$\OO_\alpha$ according to the form degree on the $M$ factor, one can write
\[
\OO_\alpha \ = \ \OO_\alpha^{(0)} \ +\ \cdots \ +\ \OO_\alpha^{(2m)}\ ,  
\]
where the restriction of $\OO_\alpha^{(j)}$ to each slice $(\varphi^a , A
,\phi )\times M$ is a $j$-form. Moreover, decomposing the identity
\[
(\dd_M + Q ) \ \OO_\alpha \ = \ 0
\]
according to the form degree on the $M$ factor, one gets the descent
equations
\begin{align*}
\dd_M \ \OO_\alpha^{2m} \ &= \ 0 \ ,  \\
\dd_M \ \OO_\alpha^{j} \ &= \ - \; Q\; \OO_\alpha^{j+1} \ , \qquad   0 \le
j \le 2m-1 \ ,  \\
0 \ &= \ Q \; \OO_\alpha^{(0)} \ .
\end{align*}
Now let $\gamma$ be any $j$-dimensional homology cycle in $M$, and define
\begin{equation}
W (\alpha , \gamma ) \ := \ \int_\gamma \ \OO_\alpha^{j}  \qquad \in
\quad  \Omega^\bullet_\GG (\AAA \times \GME ) \ .
\label{5.6}
\end{equation}
As usual, it follows from the descent equations and Stokes' theorem that $W
(\alpha, \gamma )$ is $Q$-closed, so it is an observable. Moreover, the
cohomology class of $W (\alpha , \gamma )$ in $H^\bullet_\GG (\AAA \times
\GME )$ only depends on the classes of $\alpha$ and $\gamma$ in
$H^\bullet_G (X)$ and $H_j (M)$, respectively.

\section{Observables II --- ``universal'' construction }

\subsection{The universal construction}

In the last section we saw how to associate with each equivariantly
closed form $\alpha \in \Omega^\bullet_G (X)$ another closed form
$\OO_\alpha \in  \Omega_\GG^\bullet (\AAA \times \GME \times M )$. As
we will see later, the form $\OO_\alpha$ can then be ``projected
down'' to a form in $(\AAA \times \GME )/ \GG \times M$ by, roughly
speaking, multiplying it by $e^{-I_{{\rm proj}}}$ and performing a
certain path integral. This construction corresponds to the
``quantum'' way of obtaining the topological invariants.

In this section we will describe an alternative ``universal''
construction that, also roughly speaking, associates directly with
each $\alpha$ a certain differential form on the quotient space $(\AAA
\times \GME )/ \GG \times M$. This construction corresponds to the
more traditional geometrical approach to the invariants. We will then
spend most of the time establishing a result that will later allow us to
relate these two constructions. (This result is formula (\ref{6.3}),
and if the reader is willing to accept it, the subsections 6.2 and 6.3
can be skipped.)

$\ $

Besides acting on $\AAA \times \GME$, the group of gauge transformations
$\GG$ also acts on the principal bundle $P$. This action is effective and
commutes with the natural $G$-action on $P$. Thus there is a natural
action of the group $\GG \times G$ on the product space $\AAA \times \GME
\times P$. Now let $\VVV$ be any $\GG$-invariant open subset or
submanifold of $\AAA \times \GME$ where $\GG$ acts freely. Then the action
of $\GG \times G$ on $\VVV \times P$ has no fixed points, and in the
commutative diagram
\begin{equation}
\begin{CD}
\VVV \times P        @>{\pi_3}>>      (\VVV \times P ) / \GG     \\ 
@V{\pi_1}VV                                     @VV{\pi_4}V
\\
\VVV \times M       @>>{\pi_2}>       \VVV /\GG  \; \times \;  M
\end{CD}
\label{6.1}
\end{equation}
all the quotient maps are principal bundles. More specifically, $\pi_1$
and $\pi_4$ are $G$-bundles, whereas $\pi_3$ and $\pi_2$ are
$\GG$-bundles. We will see later that there are natural connection forms
$\theta \in \Omega^1 (\VVV \times M ;  \Gg )$ on the bundle $\pi_2$ and
$\beta \in \Omega^1 ((\VVV \times P) / \GG ; \g )$ on the bundle $\pi_4$.

$\ $
  
At this point recall the evaluation map $\Upsilon : \GME \times M
\rightarrow E$ defined in (\ref{3.15}). Since elements of $\GME$ can be
identified with $G$-equivariant maps $P \rightarrow X$, the evaluation map
$\Upsilon$ can be identified with a map $\GME \times P \rightarrow X $,
and this can be trivially extended to
\begin{equation*}
\tilde{\Upsilon} : \ \AAA \times \GME \times P \ \longrightarrow \ X \ .
\end{equation*}
It follows straightforwardly from the definitions that $\tilde{\Upsilon}$
is $G$-equivariant and is constant on the $\GG$-orbits in $\AAA \times   
\GME \times P$. Thus, restricting to $\VVV$, $\tilde{\Upsilon}$ induces a
$G$-equivariant map
\[
\hat{\Upsilon} : \ (\VVV \times P )/\GG \ \longrightarrow \ X .
\]
Hence given any equivariantly closed form $\alpha \in \Omega^\bullet_G
(X)$, we get by pull-back another equivariantly closed form
$\hat{\Upsilon}^\ast \alpha  \ \in \ \Omega^\bullet_G ((\VVV \times P) /
\GG )$.

Now it is true on general grounds that the equivariant cohomology of the
total space of a principal bundle is isomorphic to the de Rham cohomology
of the base space of the bundle. An explicit isomorphism may be
constructed by choosing a connection on the bundle and applying the Weil
homomorphism \cite{B-G-V, C-M-R}. In our present problem, we can use
the connections 
$\beta$ and $\theta$ to define Weil homomorphisms
\begin{align}
w_\beta &: \ \Omega^\bullet_\GG ((\VVV \times P) / \GG) \ \longrightarrow
\ \Omega^\bullet ((\VVV \times P) / \GG )_{G-{\rm basic}} \ \simeq \
\Omega^\bullet (\VVV / \GG  \times M )    \label{6.2}   \\
w_\theta  &: \ \Omega^\bullet_\GG ( \VVV \times M ) \ \longrightarrow \
\Omega^\bullet ( \VVV \times M )_{\GG-{\rm basic}} \ \simeq \
\Omega^\bullet (\VVV / \GG  \times M ) \ .   \nonumber
\end{align}
The aim of this section is to show that
\begin{equation} 
w_\beta ( \hat{\Upsilon}^\ast \alpha ) \ = \ w_\theta (\OO_\alpha )
\label{6.3}
\end{equation} 
as differential forms on the moduli space $\VVV / \GG \times M $. This   
result is important for the identification of the invariants obtained by 
quantum field theory methods, with the invariants obtained by more
traditional geometrical approaches.
\vspace{.3cm}
\begin{rem}
In  Section 7 we will take $\VVV$ to be the space of solutions of the
vortex equations, and it is not always true that $\GG$ acts freely on this
space. In fact, in the special case of pure Yang-Mills, this never
happens, and there one is forced to work with framed connections and deal
with the reducible instantons. In the case of our gauged sigma-model one
can hope that in some instances this problem will be less acute. This is
because a gauge transformation that preserves the connection $A$ in $(A,
\phi) \in \AAA \times \GME$ may not preserve the section $\phi$, and so 
the $\GG$-stabilizers will in general be ``smaller''. This is
confirmed in some examples in Section 7.2. This problem
nevertheless still requires a more careful study.
\end{rem}

\subsection{The natural connections $\theta$ and $\beta$}

The purpose of this subsection is to describe the natural connection
forms $\theta$ and $\beta$ mentioned in the discussion above. We will
also give some formulae for the curvature forms of these
connections. The presentation is rather summarized, and most of the
calculations are omitted.

$\ $

We will start with the connection $\theta$. As described in Section
3, the right $\GG$-action on $\VVV$ induces operators
\begin{equation}
C_{(A , \phi )} \ =\ C_A + C_\phi : \ \Gg \ \longrightarrow \
T_{(A , \phi )} \VVV \ . 
\label{6.301}
\end{equation}
Using the $\GG$-invariant metrics on $\VVV$ and $\Gg$, one then
defines the adjoints
\begin{equation}
C^{\dag}_{(A , \phi )} \ =\ C^{\dag}_A + C^{\dag}_\phi : \
T_{(A , \phi )} \VVV \ \longrightarrow \Gg \ , 
\label{6.31}
\end{equation}
and $C^\dag$ can be regarded as a $1$-form on $\VVV$ with values in
$\Gg$. Since the action of $\GG$ on $\VVV$ is free, the maps $C_{(A
  , \phi )}$ are injective. Moreover, since the kernel of
$C^\dag_{(A , \phi )}$ is the orthogonal complement to the image of
$C_{(A , \phi )}$, the linear map
\begin{equation}
C^{\dag}_{(A , \phi )}  C_{(A , \phi )} : \Gg \ \longrightarrow
\ C^{\dag}_{(A , \phi )} (T_{(A , \phi )} \VVV ) 
\label{6.32}
\end{equation}  
is an isomorphism. One can therefore define a $\Gg$-valued form
$\theta$ on $\VVV$ by the formula 
\[
\theta_{(A , \phi )} \ =\ \left( C^{\dag}_{(A , \phi )}
C_{(A , \phi )} \right)^{-1} \circ C^{\dag}_{(A , \phi )} \ .
\]
The $\GG$-equivariance of this form follows from the $\GG$-invariance
of the metrics on $\VVV$ and $\Gg$. Since it is also clear that
$\theta_{(A ,\phi )} \circ C_{(A , \phi )} = {\rm id}_{\Gg}$,
one concludes that $\theta$ is a connection form for the bundle $\VVV
\rightarrow \VVV / \GG $. This form can be trivially extended to a
$\Gg$-valued form on the product $\VVV \times M$, which we also call
$\theta$. This extension is a connection form for the bundle $\pi_2 :
\VVV \times M \rightarrow \VVV / \GG \times M$.

Now we denote by ${\mathcal H}_\theta$ and ${\mathcal F}$,
respectivelly, the horizontal distribution and the curvature form on
$\VVV$ determined by the connection $\theta$. It is clear from the
definition of $\theta$ that ${\mathcal H}_\theta$ is just the
orthogonal complement in $T\VVV$ to the image of $C_{(A , \phi
  )}$. As for the curvature ${\mathcal F}$, which is a $\Gg$-valued
$2$-form on $\VVV$, one can compute that 
\begin{eqnarray*}
{\mathcal F}_{(A , \phi )} ( a_1 + V_1\, ,\, a_2 + V_2) \ &= & \
(\tilde{\dd} \theta )_{(A , \phi )} \; ( a_1 + V_1\, ,\, a_2 + V_2
) \ =  \\
& = & \ -\, 2 \, (C^\dag C )^{-1} \; \{ \; [e^2\, \kappa^{ab}\,
  (g_X)_{ts}\, V_1^r \,  V^s_2 \,(\nabla \he_b )^t_r \; + \\    
&   & \  + (g_M)^{\mu \nu }\, (a_1)^c_\nu \, (a_2)^d_\mu \, f_{cd}^a
  \; ] \ e_a    \}  
\end{eqnarray*} 
for any horizontal vectors $a_i +V_i \; \in \; {\mathcal H}_\theta \subset
T_{(A , \phi )} \VVV  \subset T_A \AAA \oplus T_\phi \GME $. The
vertical vectors in $T\VVV$ are of course annihilated by ${\mathcal
  F}$.

$\ $

Having dealt with $\theta$, we now describe the connection $\beta$ on
the bundle $\pi_4$. For this we start by recalling the injective
linear map
\[
I : \Omega^k (M ; \g_P) \longrightarrow \Omega^k (P ; \g ) 
\]
determined by the formula
\[
(\pi^\ast_P \, \nu )_p \; (Y_1, \ldots , Y_k ) \ =\ [\; p \; , \; I(\nu)
  \; (Y_1 , \ldots , Y_k ) \; ]  \qquad \in \ (\g_P )_{\pi_P (p) } \ .
\]
Here $\nu$ is any form in $\Omega^k (M ; \g_P )$, $p$ is any point in
$P$, and $Y_j$ is any vector in $T_p P$. The image of this map is
exactly the set of $G$-equivariant horizontal forms in $\Omega^k (P ;
\g )$. In other words, it is the set of forms $\alpha \in \Omega^k (P
; \g)$ that satisfy $R_h^\ast \alpha = {\rm Ad}_h \circ \alpha$ and
$\iota_Y \alpha = 0$ for all $h$ in $G$ and all $Y$ in $\ker (\dd
\pi_P)$. The map $I$ and the connection form $\theta \in  \Omega^1
(\VVV  ; \Gg )$ allow us to define a form $I \circ \theta \in
\Omega^1 (\VVV \times P ; \g )$ by the formula 
\[
(I \circ \theta )_{(A , \phi , p )} (a+ V + Y ) \ =\ I [\: \theta_{(A,
    \phi )} (a+V )\: ]\  |_p  \qquad \in \ \g \ ,
\]
where $a+ V$ is any vector in $T_{(A , \phi )} \VVV \; \subset \; T_A
\AAA \oplus T_\phi \GME $ and $Y$ is any vector in $T_p P$.

Besides $I \circ \theta$, there is another natural form in $\Omega^1
(\VVV \times P ; \g ) $, which is actually a connection form on the
bundle $\pi_1$. This is the form $\eta$ defined by the formula 
\begin{equation}
\eta_{(A , \phi , p)} (a+V+Y) \ =\ A_p (Y) \qquad  \in \ \g \ .
\label{6.4}
\end{equation}
Thus we can form the combination $\eta + I\circ \theta $, which is a
$\g$-valued $1$-form on the manifold $\VVV \times P$. A more careful
study of this form, which we omit here, then shows that $\eta + I\circ
\theta $ descends to a form on the quotient $(\VVV \times P ) / \GG $,
i.e. 
\begin{equation}
\eta \; + \; I\circ \theta \ =\ \pi_3^\ast \; \beta 
\label{6.5}
\end{equation}
for a unique $\beta \in \Omega^1 ((\VVV \times P )/ \GG ; \g
)$. Moreover, one can also show that this natural form $\beta$ defines
a connection on the bundle $\pi_4$, as desired.

$\ $

Now denote by $F_\beta$ the curvature of the connection $\beta$, which
is an element of $\Omega^2 ((\VVV \times P)/\GG ; \g )$. A computation
using (\ref{6.5}) shows that
\begin{align}
&(\pi_3^\ast \; F_\beta )_{(A , \phi , p)} (Y_1 , Y_2) \ =\ (F_A)_p (Y_1
  , Y_2)        \label{6.6}     \\
&(\pi_3^\ast \; F_\beta )_{(A , \phi , p)} (a_1 +V_1 , a_2 +V_2 ) \ =\ I
  (\; {\mathcal F}_{(A , \phi )} ( a_1 + V_1\, ,\, a_2 + V_2)  \;
  )\ |_p      \nonumber \\
&(\pi_3^\ast \; F_\beta )_{(A , \phi , p)} (a+ V , Y ) \ =\ I(a)_p \
  [Y]  \nonumber 
\end{align}
for any horizontal vectors $a_j + V_j \in {\mathcal H}_\theta  \subset
T_{(A , \phi)} \VVV $ and $Y_j \in {\mathcal H}_A \subset T_p P$.

What we really need for the next subsection, however, is the pull-back
of $\pi_3^\ast \, F_\beta $ by any local section of the bundle
$\pi_1$, and this is what we will now compute. Let $s: \UU \rightarrow
P$ be a local section of $P$ over a domain $\UU \subset M$. It determines
a local frame of $\g_P$ by 
\[
x \ \longmapsto \ e_b (x) := [s(x), e_b] \ \ \in \ (\g_P)_x 
\]
for all $x \in \UU$ and $e_b$ in a basis of $\g$. Any form $\nu$ in
$\Omega^k (M ; \g_P )$ can then be locally written as $\nu = \nu^b (x)
\ e_b (x)$, and it follows from the definition of the map $I$ that
\begin{equation}
\nu^b \ =\ s^\ast \: I(\nu)^b  \qquad {\rm for\ all} \ \ b=1, \ldots ,
\dim \g \ .
\label{6.7}
\end{equation}
In particular we have that, for any $v \in T_x \UU$ and any $a \in
\Omega^1 (M ; \g_P )$,
\[
I(a)_{s(x)} \ [\: (\dd s)_x (v)\: ] \ =\ a^b_x (v) \ e_b (x) \ =\ (\psi_\mu^b
\wedge \dd x^\mu )_{(A, \phi , x)} (a, v) \ e_b (x) \ ,
\]
where in the last term we regard $a$ as an element of $T_A \AAA$. Thus
considering (\ref{6.6}), (\ref{6.7}) and the fact that $F_A$ and ${\mathcal
  F}_{(A, \phi)}$ are horizontal forms, we get that
\begin{align}
&(s^\ast \pi_3^\ast \; F_\beta )_{(A , \phi , x)} (v_1 , v_2) \ =\
  (s^\ast F_A)_x (v_1 , v_2)   \label{6.8}  \\
&(s^\ast \pi_3^\ast \; F_\beta )_{(A , \phi , x)} (a_1 +V_1 , a_2 +V_2 ) \ =\ 
   {\mathcal F}_{(A , \phi )} ( a_1 + V_1\, ,\, a_2 + V_2)  
  \ |_x   \nonumber    \\
&(s^\ast \pi_3^\ast \; F_\beta )_{(A , \phi , x)} (a+V , v) \ =\
  (\psi_\mu \wedge \dd x^\mu )_{(A, \phi , x)}^{{\mathcal H}_\theta
  \oplus TM} (a+V , v )         \nonumber
\end{align}
for any $a_j + V_j \in T_{(A , \phi)} \VVV$ and $v_j \in T_x M$. In
this formula the symbol
\[
(\psi_\mu \wedge \dd x^\mu )^{{\mathcal H}_\theta
  \oplus TM}
\]
denotes the composition of the form $\psi_\mu \wedge \dd x^\mu$ on
$\VVV \times M$ with the projection of vectors
\begin{equation}
T(\VVV \times M ) \ =\ \ker (\dd \pi_2) \oplus {\mathcal H}_\theta
 \oplus TM \ \longrightarrow {\mathcal H}_\theta \oplus TM \ ,
\label{6.81}
\end{equation}
i.e. the horizontal part of $\psi_\mu \wedge \dd x^\mu$ with respect
to the connection $\theta$ on the bundle $\pi_2$. By an abuse of
notation, we have also used the same symbol for the section $s$ and its
trivial extension $s : \VVV \times \UU \rightarrow \VVV \times P$. The
forms $s^\ast \pi_3^\ast F_\beta \in \Omega^2 (\VVV \times \UU ; \g)$
that we have just calculated will be essential in the next subsection
to establish the identity (\ref{6.3}).

\subsection{Comparing the two constructions}

The aim of this subsection is to justify equality (\ref{6.3}). This
equality is the fundamental relation between the ``quantum'' approach
of Section 5 and the universal construction of 6.1.

$\ $

Start
by considering the Weil homomorphisms (\ref{6.2}), and regard the forms
$w_\beta (\hat{\Upsilon}^\ast \alpha )$ and $w_\theta (\OO_\alpha) $
as basic forms on $(\VVV \times P) / \GG $ and $\VVV \times M$,
respectivelly. Since both these forms descend to $\VVV / \GG \times
M$, the commutativity of diagram (\ref{6.1}) implies that (\ref{6.3}) is
equivalent to 
\[
\pi_3^\ast \ w_\beta (\hat{\Upsilon}^\ast \alpha ) \ = \ \pi_1^\ast \
w_\theta (\OO_\alpha)  \ .
\]
It is enough to establish this identity locally, so all we have to do
is to show that
\[
w_\theta (\OO_\alpha) \ = \ s^\ast \pi_3^\ast \ w_\beta
(\hat{\Upsilon}^\ast \alpha )  \ ,
\] 
where $s : \VVV \times \UU \rightarrow \VVV \times P$ is the local
trivialization of $\pi_1$ described at the end of the last subsection.

Now let the equivariant form $\alpha$ be as in (\ref{5.3}), so that
\[
\hat{\Upsilon}^\ast \alpha \ = \ \frac{1}{k!\, l!} \; (\, \alpha_{a_1 \cdots
  a_k r_1 \cdots r_l } \circ \hat{\Upsilon}\, ) \ 
\zeta^{a_1} \cdots \zeta^{a_k} \; (\hat{\Upsilon}^\ast \dd u^{r_1})
  \wedge \cdots \wedge  ( \hat{\Upsilon}^\ast \dd u^{r_l}) \ .
\]
By definition of the Weil homomorphism, this form is taken to
\begin{equation}
w_\beta (\hat{\Upsilon}^\ast \alpha ) \ = \ \frac{1}{k!\, l!} \;
 (\, \alpha_{a_1 \cdots a_k r_1 \cdots r_l } \circ \hat{\Upsilon}\, ) \ 
F^{a_1}_\beta \wedge  \cdots \wedge F^{a_k}_\beta  \wedge
  (\hat{\Upsilon}^\ast \dd u^{r_1})_{{\rm hor}} 
  \wedge \cdots \wedge  ( \hat{\Upsilon}^\ast \dd u^{r_l})_{{\rm hor}} \ ,
\label{6.82}
\end{equation}
which is a $G$-basic form on $(\VVV \times P) / \GG$. Here the
subscript ``hor'' means the horizontal part of the forms with respect
to the connection $\beta$.

$\ $

Now in general, for any form $\nu$ in $\Omega^1 (X)$, we have that
\begin{equation*}
( \hat{\Upsilon}^\ast  \nu )_{{\rm hor}} \ =\ \hat{\Upsilon}^\ast \nu
\ -\ \beta^a \; (\hat{\Upsilon}^\ast \nu ) [\check{e}_a] \ ,
\label{6.9}
\end{equation*}
where $\check{e}_a$ is the vector field on $(\VVV \times P) / \GG$
associated to $e_a \in \g$ by the right action of $G$. But from the
$G$-equivariance of the map $\hat{\Upsilon}$ it is also clear that
\begin{equation*}
(\dd \hat{\Upsilon} ) (\check{e}_a ) \ =\ - (\he_a) \ ,
\end{equation*}
where $\he_a$, as usual, is the vector field on $X$ associated to $e_a
\in \g$ by the left $G$-action on $X$. Hence one obtains that
\[
(\hat{\Upsilon}^\ast \nu )_{{\rm hor}} \ =\ \hat{\Upsilon}^\ast \nu \;
+ \; \beta^a \; (\iota_{\he_a} \nu ) \circ \hat{\Upsilon} \ ,
\]
and therefore
\begin{equation}
\pi_3^\ast \: (\hat{\Upsilon}^\ast \nu )_{{\rm hor}} \ =\
\tilde{\Upsilon}^\ast \nu \; + \; (\pi_3^\ast \beta^a ) \;
(\iota_{\he_a} \nu ) \circ \tilde{\Upsilon} \ . 
\label{6.10}
\end{equation}
On the other hand, it follows from (\ref{6.5}), (\ref{6.4}) and (\ref{6.7})
that 
\[
s^\ast \, \pi_3^\ast \: \beta^a \ |_{(A, \phi , x)} \ = \ (s^\ast A^a )_x \;
+ \; \theta^a_{(A , \phi )} \ .
\]
Moreover, tautologically,
\begin{equation}
\tilde{\Upsilon} \circ s \; (A, \phi , x) \ =\ \hphi (x) \ ,
\label{6.11}
\end{equation}
where $\hphi : \UU \rightarrow X$ is the local representative of $\phi
\in \GME$ with respect to the trivialization of $E$ induced by
$s$. Thus the pull-back by $s$ of equation (\ref{6.10}) is 
\begin{equation}
s^\ast \, \pi_3^\ast \: (\hat{\Upsilon}^\ast \nu )_{{\rm hor}} \ |_{(A ,
  \phi , x )}\ =\
s^\ast \, \tilde{\Upsilon}^\ast \nu \ + \ (\iota_{\he_a} \nu ) \circ \hphi \
 [ s^\ast A^a + \theta^a ] \ ,
\label{6.12}
\end{equation}
where  on the right hand side we have omitted the dependence on $x \in
M$. 

$\ $

The equation above will now be applied to the particular case where
$\nu$ is the local $1$-form $\dd u^r$. Denoting by $\tilde{\dd}$ the
exterior derivative on $\AAA \times \GME \times M$, and noting that
$\tilde{\dd }$ coincides with the equivariant differential $\dd_M + Q$
when acting on functions, it follows from (\ref{6.11}) that
\[
s^\ast \: \tilde{\Upsilon}^\ast \; \dd u^r \ |_{(A , \phi , x )} \ =\
\tilde{\dd} [\hphi^r (x)] \ =\ (\dd \hphi^r )_x \ + \ \chi^r (x) \;
|_{\phi} \ .
\]
On the other hand, considering the horizontal projection (\ref{6.81}), one
can compute that the component in ${\mathcal H}_\theta \oplus TM$ of the
$1$-form $\chi^r (x) \in T^\ast_{(\AAA , \phi , x )} (\VVV \times M)$
is given by
\begin{align*}
[\chi^r (x)]^{{\mathcal H}_\theta \oplus TM} (a + V + v) \ &= \ \chi^r
(x) \; \big[\: a+V+v \, - \, C_{(A , \phi)} \circ \theta (a+V+v) \:  \big] \ = \\
&=\ [\: \chi^r (x) \; + \; \he_b^r \circ \hphi (x) \, \theta^b \: ] \;
(a+V+v) \ .
\end{align*}
Hence it follows from (\ref{6.12}) that
\begin{equation}
s^\ast \, \pi_3^\ast \: (\hat{\Upsilon}^\ast \: \dd u^r )_{{\rm hor}} \ |_{(A ,
  \phi , x )}\ =\ (\dd^A \hphi^r )_x \ + \ [ \chi^r (x) ]_{(A,\phi ,
  x)}^{{\mathcal H}_\theta \oplus TM } \ .
\label{6.13}
\end{equation}
With the formula above at hand, it is now possible to compute the
pull-back by $\pi_3 \circ s$ of equation (\ref{6.82}). In fact, making
use of (\ref{6.8}) and (\ref{6.13}), we have that
\begin{align}
s^\ast \, \pi_3^\ast \: w_\beta ( \hat{\Upsilon}^\ast \alpha ) \ |_{(A ,
  \phi , x )}\ =& \frac{1}{k!\, l!} \; \alpha_{a_1 \cdots a_k r_1 \cdots
  r_l } \circ \hphi (x)        \label{6.14}  \\
&\ \ \Big[ \ \prod_{1\le j\le k } \; ({\mathcal F}^{a_j} + \psi_\mu^{a_j}
  \wedge \dd x^\mu  + F^{a_j} )_{(A , \phi , x)} \ 
  \Big]^{{\mathcal H}_\theta \oplus TM  }  \wedge    \nonumber \\
& \ \ \wedge \Big[ \ \prod_{1\le i\le l } \; ( \chi^{r_i} + \dd^A \hphi^{r_i}
  )_{(A , \phi , x)}\  \Big]^{{\mathcal H}_\theta \oplus TM  } \ ,
  \nonumber 
\end{align}
where we have also used that ${\mathcal F}^{a}_{(A, \phi)}$,  $(F_A^a
)_x$ and  $(\dd^A \hphi^r )_x$, regarded as forms in
$\Lambda^\bullet_{(A , \phi , x)} (\VVV \times M)$, are already
horizontal with respect to the projection $T(\VVV \times M)
\rightarrow {\mathcal H}_\theta \oplus TM $.

The final step is to compare the expression above with the definition
of the equivariant form $\OO_\alpha$. It is then clear that (\ref{6.14})
can be obtained from (\ref{5.4}) by substituting $\varphi^a \rightarrow
{\mathcal F}^a $ and taking the ${\mathcal H}_\theta \oplus
TM$-component of the resulting form. But as is well known, this is
precisely the definition of the Weil homomorphism
\[
w_\theta : \ \Omega_\GG^\bullet (\VVV \times M ) \ \longrightarrow \
\Omega^\bullet (\VVV \times M )_{\GG {\rm -basic}}  
\]
associated with the connection $\theta$ on the bundle $\VVV \times M
\rightarrow \VVV /\GG \times M$. Thus the right hand side of (\ref{6.3})
coincides with $w_\theta (\OO_\alpha)$, as desired.

\section{Invariants and localization}

\subsection{Correlation functions and localization}

The purpose of this final section is to study the correlation
functions of the observables $W (\alpha , \gamma )$ defined in Section
5. As is usual in topological field theory, the importance of these
correlation functions stems from the fact that they are expected to be
invariant under deformations of the metric and complex structure of
the manifolds $M$ and $X$. This means that they essentially only
depend on the $G$-action and on the differentiable/symplectic structures of $M$
and $X$, and hence are potentially able to distinguish inequivalent
manifolds and $G$-actions. Another important property of the
correlation functions is that, while they are defined by a certain
path-integral over the space of all fields, their computation can be
reduced to an integral over the moduli space of
solutions of the vortex equations.

We point out that the methods of this first subsection are
standard, as the localization arguments that apply to topological
Yang-Mills, for instance, can be straightforwardly transposed to our 
gauged sigma-model; at least at a heuristical level. Thus, besides the
original references \cite{W1, W2}, we 
follow closely the review \cite[ch.14]{C-M-R}.

$\ $

The correlation functions of the observables $W(\alpha , \gamma )$ are
of the form
\begin{equation}
Z (\alpha_1 , \gamma_1 , \ldots , \alpha_k , \gamma_k ) \ := \ \int
{\mathcal D} (b,c,d,\varphi , \eta , \lambda , A , \psi , \phi , \chi
) \ \ e^{-I_{{\rm loc}} -I_{{\rm proj}}} \ \prod_i \: W(\alpha_i , \gamma_i ) \ .
\label{7.1}
\end{equation}
Since the path-integral measure is assumed not to depend on the metric
and complex structure on $M$ and $X$, the dependence of $Z$ on these
quantities is contained in 
\[
I_{{\rm loc}} + I_{{\rm proj}} \ =\ Q \: (\Psi_{{\rm loc}} +
\Psi_{{\rm proj}} ) \ . 
\]
Therefore under a small deformation $\delta g_{\mu \nu}$ of the metric
on $M$, for example, the change $\delta Z$ is given by the
path-integral of  
\[
e^{-I_{{\rm loc}} -I_{{\rm proj}}} \ \ Q \: \Big[ \: (\delta \Psi_{{\rm loc}} +
\delta \Psi_{{\rm proj}} ) \; \prod_i \: W(\alpha_i , \gamma_i ) \: \Big]
\ ,
\]
where we have used that the $W's$ are $Q$-closed and that the
differential $Q$ does not depend on $g_{\mu \nu}$ (see the remarks
after (\ref{4.2})). This last path-integral represents the vacuum expectation
value of a $Q$-exact quantity, and by standard heuristical arguments
it vanishes \cite{W1, W2}. One therefore expects the correlation functions
$Z$ to be invariant under deformations of the metric and complex
structure of $M$ and $X$ (see however Section 7.3).

$\ $

On the other hand there is the localization argument, which reduces
the path-integral defining $Z$ to an integral over the moduli space of
vortex solutions. In order to state this result, let $\VVV$ denote the
space of solutions of the vortex equations, and assume that $\GG$ acts
freely on $\VVV$. The basic localization result, as stated in \cite{C-M-R},
asserts that
\begin{align}
Z\ &= \ \int_{(A, \phi) \in \VVV} {\mathcal D}(A, \phi , \psi , \chi ,
\varphi , \eta , \lambda ) \ \ \Big( \prod_i W_i \Big)\; \wedge\;
	{\mathcal E}({\rm 
 cok}\: \OOO  \rightarrow \VVV ) \ \ e^{-I_{{\rm proj}}}   \nonumber  \\
&= \ \int_{\VVV /\GG } \Big[  \prod_i  w_\theta (W_i ) \Big]\;
\wedge\;  {\mathcal E}({\rm cok}\: \OOO /\GG  \rightarrow \VVV /\GG ) \ .  
\label{7.2}
\end{align}
We now explain the notation in this formula. $\OOO$ is the linear
operator defined at each $(A ,\phi ) \in \AAA \times \GME $ by 
\begin{equation}
\OOO_{(A , \phi)}\; = \; (\tilde{\nabla} s )_{(A , \phi )} \oplus
\frac{1}{\sqrt{2} e}\, C^\dag_{(A ,\phi )} \: : \ T_{(A, \phi )} (\AAA
\times \GME) \ \longrightarrow  \ \VV_{(A , \phi )} \oplus \Gg \ ,
\label{7.3}
\end{equation}
where we use the notation of Appendix A. At each vortex solution $(A
,\phi ) \in \VVV $, identifying the target of $\OOO$ with the space
\[
\Omega^{0,1} (M ; \phi^\ast \ver )\: \oplus \: \Omega^{0,2} (M ; (\g_P
)_{\mathbb C} )\:  \oplus \: \Omega^0_+ (M ; \g_P )\: \oplus\:  \Omega^0 (M ;
\g_P ) \ , 
\]
a calculation shows that this operator can be written in local
coordinates as
\begin{equation}
\OOO_{(A ,\phi)} \ =\ 
\left[  \begin{array}{c}
\left\{\: [\, (\phi^\ast \nabla^A)^{0,1} \chi ]^j_{\bar{\alpha}}\: +\: 
\psi^a_{\bar{\alpha}}\, \he^j_a\:  \right\} \;  \dd \bar{z}^\alpha \otimes
\hphi^\ast (\frac{\partial}{\partial w^j })   \\
\frac{1}{\sqrt{2} e}\, \left\{\: \overline{(D_A \psi)^a_{\alpha \beta}}\: -
\: \overline{(D_A \psi)^a_{\beta \alpha }} \:  \right\} \  (\dd
\bar{z}^\alpha  \wedge \dd \bar{z}^\beta ) \ e_a   \\
\frac{1}{\sqrt{2} e}\, \Im {\mathfrak m} \: \left[ 4\, h^{\alpha \bar{\beta}}
\,  (D_A \psi)^a_{\alpha \bar{\beta}}\: +\: e^2\, \kappa^{ab}\, h_{j \bar{k}}\,
  \overline{\he^k_b } \, \chi^j \:  \right] \; e_a    \\
\frac{-1}{\sqrt{2} e}\, \Re {\mathfrak e} \: \left[\: 4\, h^{\alpha
    \bar{\beta}}\, 
  (D_A \psi)^a_{\alpha \bar{\beta}}\: +\: e^2 \, \kappa^{ab}\,  h_{j
    \bar{k}}\, 
  \overline{\he^k_b }\,  \chi^j \:  \right] \; e_a 
\end{array} \right] \ .
\label{7.4}
\end{equation}
The first three components of $\OOO_{(A , \phi )}$ correspond to the
operators obtained from the linearization of the three vortex equations
at the point $(A ,\phi) \in \VVV$. The last component of $\OOO_{(A ,
  \phi)}$, roughly speaking, measures the orthogonality of a tangent
vector in $T_{(A ,\phi )} (\AAA \times \GME)$ to the $\GG$-orbit of $(A
,\phi )$. The cokernels of the operators  $\OOO_{(A , \phi )}$ for all
$(A , \phi )$ in $\VVV$ define a vector bundle ${\rm cok}\: \OOO
\rightarrow \VVV $. Taking the quotient by $\GG$ one obtains another
vector bundle, and the symbol ${\mathcal E}({\rm 
 cok}\: \OOO /\GG \rightarrow \VVV /\GG )$ denotes the Euler
class of this bundle. 
Finally, the symbol $w_\theta$ represents the Weil homomorphism
(\ref{6.2}). More precisely, what we mean in formula (\ref{7.2}) is 
\[
w_\theta \: [\: W(\alpha , \gamma )\: ] \ :=\ \int_\gamma w_\theta
[\, \OO_\alpha \, ] \ . 
\]

Thus we see that the correlation functions $Z$ can be computed by
integrating certain closed differential forms over the moduli space
$\VVV / \GG$. Moreover, using the results of Section 6, we have that
\begin{equation}
Z \ =\ \int_{\VVV /\GG } \Big[  \prod_i \: \int_{\gamma_i} w_\beta \, 
	(\hat{\Upsilon}^\ast \alpha_i ) \Big] \; \wedge \; 
	{\mathcal E}({\rm cok}\: \OOO /\GG  \rightarrow \VVV /\GG ) \ .  
\label{7.5}
\end{equation}
Comparing this formula with the definition of the Hamiltonian
Gromov-Witten invariants in \cite{C-G-M-S}, one recognizes that, in the case
where ${\rm cok}\, \OOO = 0$, our correlation functions essentially
coincide with those invariants.

\subsection{The moduli space of vortex solutions}

As was seen above, the moduli space $\VVV / \GG$ is of the utmost
importance for the calculation of the correlation functions of our
theory. In this subsection we will report some properties of this
moduli space. The majority of the results here comes from references
\cite{MiR, C-G-S, C-G-M-S}. As in those references, we will restrict
ourselves to the case where $M$ is a compact Riemann surface.

$\ $

For the general gauged sigma-model the moduli space $\VVV /\GG$ is not
necessarily a smooth 
manifold. When it is a smooth manifold, or  at least in each smooth
region, the tangent space $T_{[A,\phi]} \VVV/\GG$ can be identified
with the kernel of the operator $\OOO_{(A ,\phi)}$. In the light of the discussion
below (\ref{7.4}) this identification is very natural, since a
tangent vector belongs to $\ker \OOO_{(A ,\phi)}$ exactly if it
satisfies the linearized vortex equations and is perpendicular to the
$\GG$-orbit of $(A ,\phi )$. More generally, to decide whether
$\VVV /\GG$ is or is not smooth, one should study in
detail the linearized equations and the orthogonality condition, i.e. the 
operator $\OOO$. In fact, the kernel,
cokernel and index of $\OOO$ are the relevant objects that characterize the
local structure of $\VVV /\GG$.

Following this cue, the 
first important result is the virtual dimension of the 
space $\VVV /\GG$.  This is given by the real index of the Fredholm
operator $\OOO$, and coincides with the actual dimension of $\VVV / \GG$
on smooth regions. 
Notice that, for $M$ a Riemann surface, the second
component of this operator in (\ref{7.4}) should be discarded, as the
vortex equation $F_A^{0,2} =0$ is trivially satisfied. The computation
of the index of $\OOO$ was performed in references \cite{C-G-S, MiR}, and the
result is 
\begin{align*}
{\rm ind}\, \OOO_{(A ,\phi)} \ &=\ \dim \ker \OOO_{(A, \phi)} \ -\
\dim {\rm cok}\, \OOO_{(A, \phi )}    \nonumber  \\
&=\ (\dim_{\mathbb C} X - \dim G )\, (2-2g) \ +\ 2\, \langle \: c_1^G (TX) ,
     [\phi] \: \rangle  \ . 
%% \label{7.6}
\end{align*}
In this formula $g$ is the genus of $M$, $c_1^G (TX)$ is the
equivariant first Chern class of $TX$ --- which belongs to $H^2_G (X ;
{\mathbb Z})$ --- and $[\phi]$ is the class in $H_2^G (X ; {\mathbb
  Z})$ determined by the section $\phi$. In practice we have that
$\langle c_1^G (TX) , [\phi] \rangle$ equals the first Chern number of
the bundle $\phi^\ast \ver \rightarrow M$. 

Observe that different connected components of the moduli space $\VVV
/ \GG$ may have different dimensions, depending on the class $[\phi
]$. In fact, in formulas such as (\ref{7.1}) and (\ref{7.2}), one usually
fixes a class $B \in H^G_2 (M ; {\mathbb Z})$, and then only
integrates over the fields $\phi$ such that $[\phi]=B$. In the case
of the vortex solutions this defines a subset $\VVV_B \subset \VVV$
which can be shown to be $\GG$-invariant.

$\ $

Still regarding the smoothness of the moduli space, the best one can 
usually do is to guarantee this smoothness on the 
(typically open) subset $\VVV^\ast \subset \VVV$ of the so-called 
irreducible solutions. In fact, over the irreducible solutions, a sufficient
condition for this smoothness is the vanishing of the cokernel of
$\OOO_{(A ,\phi)}$, or in other words the surjectivity of this operator.
%% this is stated on page 27 of reference \cite{C-G-S}, and on page
%% 82, lemmas 3.3.1-3.3.3 of reference \cite{MiR} (thesis)
We will now give the definition of irreducible solution and state a condition
that is equivalent to the surjectivity of $\OOO_{(A ,\phi)}$. Both of these 
come from Reference \cite{C-G-M-S}.

\vspace{.2cm}

\begin{defn}
A solution $(A,\phi)$ of the vortex equations is called irreducible if
there exists a point $z\in M$ such that the stabilizer of
$\hat{\phi}(z) \in X$ is trivial, and the intersection
$\hat{\g}_{\hat{\phi}(z)} \cap J_X \: \hat{\g}_{\hat{\phi}(z)}$ is
zero. (Here $\hat{\g}_{\hat{\phi}(z)}$ denotes the image of the linear
map $\g \rightarrow T_{\hat{\phi}(z)} X$ associated to the $G$-action on
$X$.) We note that the subset $\VVV^\ast \subset \VVV$ of irreducible
vortex solutions is $\GG$-invariant, and that $\GG$ acts freely 
on it.
\end{defn}

\begin{prop} 
Consider the operator
\begin{align*}
L_{(A,\phi)}:\ \Omega^0 (M ; \phi^\ast \ver )\; \oplus \; \Omega^{0,1} (M ;
(\g_P)_{\mathbb C} ) \ &\longrightarrow \ \Omega^{0,1} (M ; \phi^\ast
\ver )    \ , \\
(\: V\: ,\ \tau_{\bar{\alpha}}^a (z) \ \dd \bar{z}^\alpha \otimes e_a
\: ) \ &\longmapsto \ (\phi^\ast \nabla^A)^{0,1}\: V \ + \
\tau_{\bar{\alpha}}^a (z) \ \dd \bar{z}^\alpha \otimes (\he_a)_{\hphi (z)}
\end{align*}
where $(\phi^\ast \nabla^A)^{0,1}$ is as in (\ref{4.101}). Then
$\OOO_{(A, \phi )}$ is surjective if and only if both $C_{(A, \phi)}$
and the adjoint $L_{(A,\phi)}^\dag$ are injective.
\label{pII.3-1}
\end{prop}

\begin{rem}
An immediate consequence of the results above is that if $L_{(A,\phi)}^\dag$
is injective for all $(A, \phi) \in \VVV^\ast$, then the moduli space 
$\VVV^\ast /\GG$ has a natural structure of smooth manifold.
\end{rem}
\vspace{.1cm}

\begin{rem}
The statement of proposition \ref{pII.3-1} is a bit stronger than 
proposition $4.8\, (iii)$ of
\cite{C-G-M-S}, but follows directly from the proof presented
there. More specifically, refer back to that proof, call $L_\phi^\CC$
the operator
\[
\Omega^1 (M ; (\g_P)_{\mathbb C} ) \ \longrightarrow\ \Omega^1 (M ;
\phi^\ast \ver )\ , \qquad \ \tau \
\mapsto \tau_\mu^a (x) \ \dd x^\mu \otimes (\he_a)_{\hphi (x)} \ ,
\]
and $L_\phi$ its restriction to $\Omega^1 (M ; \g_P )$. Then it is
enough to notice that $2 (L_\phi L_\phi^\dag \eta )^{0,1} = L_\phi^\CC
(L_\phi^\CC)^\dag \eta$ for all $\eta$ in $\Omega^{0,1} (M ; \phi^\ast
\ver )$. Moreover, $(L_\phi^\CC)^\dag \eta = 0$ if and only if
$L_\phi^\dag \eta = L_\phi^\dag (\eta \circ J_M ) =0$.
\end{rem}
\vspace{.2cm}
One of the issues raised by the previous proposition is the 
injectivity of
$C_{(A,\phi)}$, the operator in (\ref{3.3}) that represents the
infinitesimal gauge transformations. This issue is interesting for its
own sake, and also came up in the discussion below
(\ref{6.31}). In fact, notice that $C_{(A,\phi)}$ is injective if and only if
the $\GG$-stabilizer of $(A,\phi)$ is discrete. Regarding these
matters there exists the following result. 

\vspace{.3cm}
\noindent
{\it {\bf \cite{C-G-M-S}} Suppose that $M$ and $\mu^{-1} (0)$ are
  compact, and that $0$ is a regular value of $\mu$ (resp. $G$ acts
  freely on $\mu^{-1} (0)$). Then there exists a constant $K > 0$ such
  that, if $e^2 ({\rm Vol}\, M) \ge K\: T_{[\phi]}$, every solution $(A
  , \phi')$ of the vortex equations with $[\phi'] = [\phi]$ has a
  discrete (resp. trivial) $\GG$-stabilizer. }
\vspace{.3cm}

Here $[\phi] \in H_2^G (M ; {\mathbb Z})$ is the equivariant
homology class already mentioned above (which, by the way, is the same
for homotopic sections $\phi$), and $T_{[\phi]}$ is the topological
energy of (\ref{2.6}). This result means that for large enough Riemann
surfaces the group of gauge transformations acts (locally) freely on
the space of vortex solutions.

In the case of abelian actions we prove the following result in
Appendix B.

\vspace{.3cm}
\noindent
{\it 
Suppose that $G$ is a torus, that $M$ and $X$ are compact, and that
the constant $(\deg P) / (e^2 \: {\rm Vol}\, M\, )$ is a regular value of
$\mu$ (see the appendix for the definition of $\deg P$). Then every solution
of the vortex equations has a discrete
$\GG$-stabilizer. If, furthermore, the torus action on $X$ has no
non-trivial finite stabilizers, then the $\GG$-action on the set of
vortex solutions is free. }
\vspace{.3cm}

\begin{rem}
The last two propositions are true even if $\dim_\CC M >
1$. In the second one, 
the condition of compact $X$ can be very much weakened (see the remark
in Appendix B); in particular the result is still valid for linear
torus actions on ${\mathbb C}^n$.
\end{rem}
\vspace{.1cm}

\begin{rem}
Suppose that 
$G$ is a $n$-torus, that $X$ is compact of complex dimension $n$, and that the
constant $(\deg P) / (e^2\: {\rm Vol}\, M)$ lies in the interior of
the polytope $\mu (X)$. Then it is not difficult to show that the 
$\GG$-action on $\VVV = \VVV^\ast$ is free and that the operators
$L^\dag_{(A,\phi)}$  are injective for every vortex solution $(A
,\phi)$. (We omit the proof here.) According to the discussion above, this implies the
smoothness of the moduli space $\VVV /\GG$, in agreement with the
results of \cite{Ba}.
\end{rem}

\subsection{About the invariants}

\subsubsection*{Wall-crossing phenomena}

In this subsection we want to illustrate the so-called wall-crossing
phenomena for the Hamiltonian Gromov-Witten invariants \cite{C-G-S}. These
refer to the occurrences where a finite deformation of the parameters
of the topological Lagrangian leads to a change in the value of the
invariants. This is in apparent contradiction with the argument
evoked at the end of $7.1$ about the vanishing of the vacuum
expectation value of $Q$-exact operators. In fact, it will be very
clear in the example below how this argument can indeed sometimes fail. 

$\ $

For our illustrative example we will consider the case of a toric
action on $X = {\mathbb C}{\mathbb P}^n$. Defining the constant
\[
\tau_0 \ :=\ (\deg P)/(e^2 \,{\rm Vol}\, M) \qquad \in \ \g \simeq {\mathbb
  R}^n \ ,
\]
it was shown in \cite{Ba} that the moduli space of vortex solutions is a
fixed non-empty manifold $\VVV$ whenever $\tau_0$ lies in the interior
of the convex polytope $\mu (X)$. If $\tau_0$ lies  outside 
this polytope the moduli space is empty.

Now suppose that we deform the moment map $\mu$ by adding to it a
constant $\tau \in {\mathbb R}^n$. This corresponds to a deformation
of the parameters of the topological Lagrangian, and so the
heuristical arguments of Section $7.1$ would seem to imply that
that the correlation functions are invariant by this deformation. In
particular all the correlation functions should vanish, since for
$\tau$ big enough $\tau_0$ lies outside $\mu (X)$, and so the integral
in (\ref{7.2}) is over the empty set. But in Reference \cite{MiR} it was
computed that for $X = {\mathbb C}{\mathbb P}^1$ and $\tau_0 \in {\rm
  int}\, \mu (X)$ there is a non-zero Hamiltonian Gromov-Witten
invariant, so the heuristical argument must fail at some point. We
will now see explicitly how this failure comes about.

Take the gauge fermion $\Psi$ of (\ref{4.3}) and substitute $\mu$ for $\mu
+ \epsilon$, where $\epsilon$ is a small constant in ${\mathbb
  R}^n$. Then the change in $Q \Psi$ is
\[
\delta \; Q\, \Psi \ =\ \pm \frac{ie}{\sqrt{2}} \; (C , \epsilon) \ , 
\]
and so the partition function, which is the simplest invariant,
changes by 
\[
\delta Z \ =\ \mp \: \int {\mathcal D}({\rm fields}) \ \;
\frac{ie}{\sqrt{2}} \; 
(C , \epsilon) \ \; e^{- Q \: \Psi} \ .
\]
Using the explicit formula for $Q \Psi$ and integrating out the
field $C$, the corresponding equation of motion is 
\[
C\ =\ \mp\, \frac{i}{2 \sqrt{2} e t} \: ( \Lambda F_A +  e^2 \, \mu
\circ \phi ) \ , 
\]
and so one recognizes that $\delta Z$ is just the vacuum expectation value of 
\[
-\, \frac{1}{4 e t} \int_M \kappa_{ab} \; (\Lambda F_A + e^2 \: \mu \circ
 \phi)^a \; \epsilon^b \ = \ \frac{ e \, ({\rm Vol}\,M)}{4t} \ \kappa
 (v , \epsilon ) \ ,
\]
with
\[
v \ =\ \tau_0 - \frac{1}{({\rm Vol}\, M )} \int_M \mu \circ \phi \ .
\]
Now, if $\tau_0$ lies in the interior of $\mu (X)$, then the vector
$v$ may have any orientation in ${\mathbb R}^n$ as $\phi$ varies, and
so it is certainly possible that the expectation value of $\kappa (v ,
\epsilon )$ vanishes. However, when $\tau_0$ lies in the boundary or
exterior of $\mu (X)$, the vector $v$ always lies in the same
semi-space of ${\mathbb R}^n$, independently of $\phi$. Hence in this
case the expectation value of $\kappa (v , \epsilon )$ cannot vanish
for a generic infinitesimal deformation $\epsilon$, and so $\delta Z$
will not vanish.

We conclude that a finite deformation of $\mu$ may well leave the
partition function $Z$ invariant, but only as long as $\tau_0$ remains
in the interior of the polytope $\mu (X)$. When $\tau_0$ crosses the
boundary of $\mu (X)$, a jump in the value of $Z$ is expected. This
point of $\tau_0$ crossing the boundary of $\mu (X)$ corresponds as
well to a drastic change in the ground states of the theory, as the
moduli space of vortex solutions jumps from $\VVV$ to the empty
set. All this is very similar to the runaway vacua phenomena in
supersymmetric quantum mechanics \cite[ch. 12.6]{C-M-R}.

The picture that seems to arise is that, in general, the space of
parameters of the topological lagrangian (and these include metrics,
complex structures, ...) is divided into different regions by internal
walls. A small deformation of the parameters will leave the
correlation functions invariant, but when a wall is crossed the
correlation functions may jump.

\subsubsection*{Adiabatic limit and Gromov-Witten invariants}

In this final paragraph we would like to mention another interesting
property of the vortex equations, namely the correspondence between
pseudo-holomorphic curves in the symplectic quotient $X/\negmedspace / G$ and
solutions of the vortex equations in the adiabatic limit $e
\rightarrow \infty$. We assume here that $G$ acts freely in $\mu^{-1}
(0)$, so that the symplectic quotient $X/\negmedspace /G :=
\mu^{-1}(0)/G $ is in a natural way an almost K\"ahler manifold.

In the limit $e \rightarrow \infty$ the vortex equations (\ref{2.7}) for
$M$ a Riemann surface become 
\begin{equation}
\db^A \phi = 0   \qquad ; \qquad \mu \circ \phi = 0 \ .
\label{7.9}
\end{equation}
It is not difficult to show that any solution of these equations
descends to a pseudo-holomorphic map $\bar{\phi}:M \rightarrow
X/\negmedspace /G$,
and that gauge-equivalent solutions descend to the same
map. Furthermore, any pseudo-holomorphic curve $\bar{\phi}$ lifts to a
solution of (\ref{7.9}) on the bundle $P\,= \,\bar{\phi}^\ast (\mu^{-1}(0)
\rightarrow X/\negmedspace /G )$, and any two different lifts are
gauge equivalent \cite{G-S}. (In passing, the connection $A$ of the lift
is the pull-back 
by $\bar{\phi}$ of the connection $\bar{A}$ on $\mu^{-1}(0)
\rightarrow X/\negmedspace /G$ determined by the $G$-invariant metric
on $\mu^{-1} 
(0)$.) One can therefore identify the moduli space of solutions of
(\ref{7.9}) with the space of pseudo-holomorphic curves on
$X/\negmedspace /G$ such
that $P \, \simeq \, \bar{\phi}^\ast (\mu^{-1}(0) \rightarrow
X/\negmedspace /G )$.

On the other hand one would expect that for $e$ big enough there
should be some sort of close correspondence betweeen the solutions of
the vortex equations (\ref{2.7}) and the solutions of (\ref{7.9}). In
particular the Hamiltonian Gromov-Witten invariants of $X$ --- which
study the moduli space of vortex solutions --- should be able to tell
something about the Gromov-Witten invariants of $X/\negmedspace /G$
--- which study 
the space of pseudo-holomorphic curves. These  matters were studied in
detail in Reference \cite{G-S}, and under suitable conditions on $M$ and
$X$, one such relation was established. In the particular case where
$X$ is a complex vector space acted by a torus and $X/\negmedspace /G$
is a toric 
variety, a very strong correspondence had been previously established
in \cite{M-P}.

\vskip 25pt
\noindent
{\bf Acknowledgements.}
I would like to thank Prof. N. S. Manton for his encouragement, 
Professor Sir Michael Atiyah for a useful discussion, and Dr. David Tong 
for pointing out references \cite{M-P, W3}.
I am supported by \lq{\sl Funda\c{c}\~ao para a Ci\^encia e
Tecnologia}\rq, Portugal, through the research grant
SFRH/BD/4828/2001.

\appendix

\section*{Appendices}

\section{The localization bundle}

In this appendix we sketch how the gauged sigma-model can be fitted into the
geometrical method for obtaining topological Lagrangians. We will have in
mind the abstract description of this method given in \cite[ch.14]{C-M-R}, so
only the features that are particular to our model will be described here.

One starts by considering the vector bundle over the space of fields $\VV
\rightarrow \AAA \times \GME$, whose fibre at a point $(A, \phi)$ of the
base is
\begin{equation*}
\VV_{(A, \phi)} \ = \ \Omega^{0,1} (M; \phi^\ast \ver ) \: \oplus \: 
\Omega_+^0 (M; \g_P ) \:  \oplus \: \Omega^{0,2}(M; (\g_P)_{\mathbb C} ) \ .
\tag{A1}
\label{a1}
\end{equation*}
In this formula $\Omega_+^0 (M; \g_P )$ is the subspace of $\Omega^0 (M;
\g_P )$ defined by
\[
\Omega_+^0 (M; \g_P ) \ := \ \Lambda ( \Omega^2 (M; \g_P ) ) \ + \
\Omega^0 (M; (\g_0)_P ) \ ,
\]
where $\Lambda$ is the operator of  (\ref{2.8}),  and $(\g_0)_P$ is the
sub-bundle of  $\g_P$ constructed from the ${\rm Ad}_G$-invariant subspace
\[
\g_0 \ := \ {\rm span}\{  \mu (X) \} \ \subseteq \ \g .
\]
We note in passing that, in the case where $X$ is a point and
$\dim_{\mathbb R} M = 4$, the space $\VV_{(A, \phi)}$ defined above is
isomorphic to the space of anti-self-dual $\g_P$-valued forms on $M$. This
is important because we want our construction to contain topological
Yang-Mills theory as a special case.

The vector bundle $\VV$ has a natural section defined by
\begin{equation*}
s\, (A, \phi) \ = \ \left( \;  \db^A \phi \; , \; \frac{1}{\sqrt{2} e}   
(\Lambda F_A + e^2 \, \mu \circ \phi ) \; , \; \frac{\sqrt{2}}{e}\, 
F_A^{0,2} \;  \right) \ .
\tag{A2}
\label{a2}
\end{equation*}
Notice that the zero set of $s$ is the set of solutions of the vortex
equations, and that the squared norm of $s$ is the non-topological term of
the energy functional (\ref{2.5}).

The group of gauge transformations $\GG$ has a natural right action on the
total space of the bundle $\VV$ which lifts the usual $\GG$-action on 
$\AAA \times \GME$. The section $s$ is equivariant with respect to these
actions, and so defines a section $\bar{s}$ of the quotient bundle
\[
\VV / \GG  \ \longrightarrow \ (\AAA \times \GME ) / \GG \ .
\]
This last bundle over the moduli space of fields is what is usually called
the localization bundle. Although our Lagrangian and observables are
ultimately meant to be defined on this bundle, it is easier to work  
"upstairs" on the bundle $\VV$, and then include a ``projection term'' that 
brings all these quantities down to the quotient bundle (see \cite{C-M-R}).

$\ $

In Section 3 the fields $A^a_\alpha (z)$ and $\hphi^j (z)$ were 
introduced as local coordinate functions on the space $\AAA \times \GME$,
which is the base of the bundle $\VV$. Here we introduce the odd fields
\[
d^j_{\bar{\alpha}} (z) \ , \quad \ \ \ c^a (z) \quad\ \ \  {\rm and}\
\ \  \quad 
b_{\bar{\alpha } \bar{\beta}}^a (z)  \ ,
\]
that should be regarded as odd coordinates on the fibre of $\VV$, or to be
more specific, respectivelly on the spaces
\[
\Omega^{0,1} (M; \phi^\ast \ver ) \ , \qquad   \Omega_+^0 (M; \g_P )
\quad \ \ \ {\rm and}\ \ \ \quad \Omega^{0,2}(M; (\g_P)_{\mathbb C} ) \ .
\]
So for example, if $\zeta = \zeta^j_{\bar{\alpha}} \: \dd \bar{z}^\alpha
\otimes \hphi^\ast (\frac{\partial}{\partial w^j})$ is an element of
$\Omega^{0,1} (M; \phi^\ast \ver )$, then the function $d^j_{\bar{\alpha}}
(z)$ evaluated at $\zeta$ gives
\[
d^j_{\bar{\alpha}} (z) \ [\zeta] \ = \ \zeta^j_{\bar{\alpha}} (z) \ .
\]

In Section 3 the operator $Q$ was defined as the differential of the
$\GG$-equivariant complex of $\AAA \times \GME$. Here we extend that
picture and define $Q$ to be the differential of the $\GG$-equivariant
complex of the total space of the bundle $\VV$. When acting on functions,
$Q$ obviously coincides with the exterior derivative $\tilde{\dd}$ on
$\VV$, so
\begin{align*}
&Q\; b_{\bar{\alpha} \bar{\beta}}^{a} (z) \ = \ \tilde{\dd} [
b_{\bar{\alpha} \bar{\beta}}^{a} (z) ]   \ =: \ B_{\bar{\alpha}
\bar{\beta}}^{a}(z)    \\
&Q\; c^a  (z) \ \, = \ \tilde{\dd} [c^a (z) ] \, \ =:  \ C^a (z)       \\
&Q\; d^j_{\bar{\alpha}} (z) \ = \ \tilde{\dd} [d^j_{\bar{\alpha}}(z)] \ =: \
D^j_{\bar{\alpha}} (z) \ -\ \Gamma_{ik}^j \circ \hphi (z) \; \chi^k (z) \;
d^i_{\bar{\alpha}}(z) \ .
\end{align*}
The rightmost equalities define the fields $B$, $C$ and $D$, which are
odd 1-forms on $\VV$. Using the explicit (local) expression for the action
of $\GG$ on $\VV$, and, as in Section 3, the definition of the
differential $Q$ of the $\GG$-equivariant complex, one can then compute
what the action of $Q$ on $B$, $C$ and $D$ is. The result is given in the
expressions (\ref{4.2}).

$\ $

The bundle $\VV \rightarrow \AAA \times \GME$ that we have been discussing
has a natural connection $\tilde{\nabla}$. This connection is trivial on  
the sub-bundles corresponding to the last two summands of (\ref{a1}), and,
on the sub-bundle corresponding to the first summand, it is naturally
induced by the Levi-Civita connection of $X$. More explicitly, let $S$ be
any section of $\VV$ and, according to (\ref{a1}), decompose it as
\[
S \ = \ S_1 + S_2 + S_3 \ ,
\]
where, locally,
\begin{align*}
S_1 (A, \phi) \ &= \  [\: (S_1 )^j_{\bar{\alpha}} (z) \: ]_{(A, \phi )} \ \dd
\bar{z}^\alpha \otimes \hphi^\ast (\frac{\partial}{\partial w^j})   \\
S_2 (A, \phi) \ &= \  [\: (S_2)^a (z)\: ]_{(A, \phi )} \ e_a  \\
S_3 (A, \phi) \ &= \  [\: (S_3 )^a_{\bar{\alpha} \bar{\beta}} (z)\: ]_{(A, \phi
)} \ ( \dd \bar{z}^\alpha  \wedge \dd \bar{z}^\beta ) \ e_a \ . 
\end{align*}
Then the connection $ \tilde{\nabla}$ acts on each of these terms as
\begin{align*}
\tilde{\nabla} S_1 \ &= \  \left\{  \tilde{\dd} [\: (S_1 )^j_{\bar{\alpha}}  
(z)\: ] \; + \;  \Gamma^j_{kl} \circ \hphi (z) \;  \chi^l (z) \; (S_1
)^k_{\bar{\alpha}} (z)   \right\}  \ \dd \bar{z}^\alpha \otimes \hphi^\ast
(\frac{\partial}{\partial w^j})   \\
\tilde{\nabla} S_2  \ &= \  \tilde{\dd} [\: (S_2)^a (z)\: ]  \ e_a  \\
\tilde{\nabla} S_3 \ &= \  \tilde{\dd} [\: (S_3 )^a_{\bar{\alpha} 
\bar{\beta}} (z)\: ]  \ ( \dd \bar{z}^\alpha  \wedge \dd \bar{z}^\beta ) \
e_a \ .
\end{align*}
Notice also that this can be rewritten as
\begin{equation*}
\tilde{\nabla} S \ = \  S^\ast [D^j_{\bar{\alpha}}] \  \dd \bar{z}^\alpha   
\otimes \hphi^\ast (\frac{\partial}{\partial w^j}) \ +\ S^\ast [C^a ]\ e_a
\ +\  S^\ast [ B^a_{\bar{\alpha} \bar{\beta}}]\  ( \dd \bar{z}^\alpha
\wedge \dd \bar{z}^\beta ) \ e_a \ .
\end{equation*}

$\ $

As mentioned in Section 4, once the field content of the theory is
established and the $Q$-action on the fields is known, there is a fairly
standard procedure to construct a $Q$-exact Lagrangian for the topological
theory. The abstract method is very well described in \cite{C-M-R},
and the necessary 
calculations are described in Section 4.

\section{A proof from Section 7}

In this appendix we want to prove the third proposition of Section
7.2. The assumptions are that $G=T^n$, $M$ and $X$ are compact, and
that the constant $(\deg P)/(e^2 \;{\rm Vol}M)$ is a regular value of
the moment map $\mu$. Here
\[
\deg P \ := \ - \int_M \Lambda F_A \qquad \in \ \g \simeq {\mathbb
  R}^n \ ,
\]
and it is not difficult to check that this constant does not depend on
the connection $A$.

For the first part of the proposition, it is enough to show that for
every vortex solution $(A ,\phi)$ the operator $C_{(A ,\phi)} = C_{A}+
C_{\phi}$ of (\ref{6.301}) is injective. So let $\varepsilon \in \Gg \simeq
C^{\infty} (M ; {\mathbb R}^n )$ be an infinitesimal gauge
transformation such that
\begin{align*}
C_{A} (\varepsilon)\ &=\ D_A \varepsilon \ = \ \dd \varepsilon \ =\ 0
\ ; \\
C_{\phi} (\varepsilon)\ &= \ - \varepsilon^a (x)\: \hphi^\ast (\he_a ) \
=\ 0 \ .
\end{align*}
The first equation tells us that $\varepsilon$ is constant. Calling
$\theta \in {\mathbb R}^n$ the constant value of $\varepsilon$, the
second equation tells us that $\hat{\theta} \: |_{\hphi (x)}
=0$, where $\hat{\theta}$ is the vector field on $X$ induced by
$\theta$.  
Using the definition of moment map, this means that $\hphi (x)
\in X$ is a critical point of the function $H_\theta := \langle \mu
,\theta \rangle $ for all $x \in M$. Now denote by $B_\theta^r$ the connected
components of the critical set ${\rm Crit} (H_\theta) \subset
X$. These components are preserved by the torus action, since $\mu$ is
$T^n$-invariant and the orbits of the action are connected. One can
therefore define the associated bundles
\[
E_\theta^r \ := \ P \times_{T^n} B_\theta^r \ ,
\] 
which are connected subsets of $E = P \times_{T^n} X$. Since $\phi (M)
\subset E$ is connected, the discussion above implies that $\phi (M)$
is contained in one of the $E_\theta^r$'s, say $E_\theta^0$. Thus
\[
\mu \circ \phi (M) \ \subset \ \mu (E_\theta^0 )= \mu (B^0_\theta )\ .
\]
Now by the lemma below, $\mu (B_\theta^0)$ is a convex polytope in
${\mathbb R}^n$, and so the constant
\[
\frac{\deg P}{e^2 \: {\rm Vol}M} \ =\ \frac{-1}{e^2 \: {\rm Vol}M}
\int_m \Lambda F_A \ =\ \frac{1}{{\rm Vol}M} \int_M \mu \circ \phi  
\]
certainly belongs to this polytope. This finally shows that, unless
$\theta =0$, the constant $(\deg P)/(e^2 \;{\rm Vol}M)$ is a critical
value of $\mu$, and the proof of the first part is complete.

For the second part of the proposition, assume furthermore that all
the $G$-stabilizers in $X$ are either trivial or a subtorus of $T^n$. To
obtain at the end a contradiction, suppose that there existed a
non-trivial gauge transformation $g \in C^{\infty} (M ; T^n)$ that
preserved the pair $(A , \phi)$. The condition that $g$ preserves the
connection $A$ implies that $g$ is a constant map, as is well
known, and we call its image also $g \in G\setminus \{ {\rm id}\}$. The
condition $g(\phi) = \phi$, on the other hand, implies that 
\[
\phi (M) \ \subset \ P \times_{T^n} {\rm Fix}(\rho_g)\ \subset E \ , 
\]
where ${\rm Fix}(\rho_g) \subset X$ is the set of fixed points of the
map $\rho_g$, and so is $T^n$-invariant. But the initial assumption
on the $G$-stabilizers in $X$ then implies that each point of ${\rm
  Fix}(\rho_g)$ is preserved by a full subtorus of $T^n$, and in
particular is a critical point of $\mu$. Thus
\[
\phi (M) \ \subset \ P \times_{T^n} {\rm Crit}(\mu)\ \subset E \ .
\]
Now, the proofs of theorem 5.47 and lemma 5.53 in \cite{MD-S} show that
\[
{\rm Crit} (\mu) \ =\ \bigcup_{\theta \in {\mathbb Z}^n \setminus \{ 0
  \}} {\rm Crit} (H_\theta )
\]
and that each ${\rm Crit} (H_\theta )$ --- the critical set of
$H_\theta$ --- is a proper complex submanifold of $X$. 
%% inspecting these proofs, the results seem to be true even for
%% noncompact X 
Thus defining
$E_\theta := P \times_{T^n} {\rm Crit} (H_\theta )$, we have that
\begin{equation}
\phi (M) \ \subset \bigcup_{\theta \in {\mathbb Z}^n \setminus \{ 0
  \}} E_\theta \ ,  \tag{B1}
\label{b.1}
\end{equation}
and that each $E_\theta$ is a proper complex submanifold of $E$
equipped with the integrable complex structure induced by $A$ (see
Section 2.2 in \cite{Ba}). On the other hand, since $\db^A \phi = 0$,
also $\phi (M)$ is a complex submanifold of $E$ (see \cite{MiR}). This
implies that the intersections $\phi (M) \cap E_\theta $ are analytic
subvarieties of $\phi (M)$, and so it follows from (\ref{b.1}) and
Baire's category theorem that there exists at least one $\theta \ne 0$
such that $\phi (M) \subset E_\theta$. Finally, arguing just as at the
end of the proof of the first part, one concludes that the constant
$(\deg P)/(e^2 \;{\rm Vol}M) $ must be a critical value of $\mu$,
which contradicts the assumptions.

\begin{lem*}
The image $\mu (B^0_\theta)$ is a convex polytope in ${\mathbb R}^n$.
\end{lem*}
\begin{proof}
Along the proof above we saw that ${\rm Crit} (H_\theta)$ and
$B^0_\theta$ are compact K\"ahler submanifolds of $X$ that are
preserved by the $T^n$-action. Thus the restriction of $\mu$ to
$B^0_\theta$ is a moment map for the $T^n$-action on $B^0_\theta$. The
lemma then follows directly from the well known convexity theorem.
\end{proof}

\begin{rem}
Inspecting the proof of the proposition presented here, it is clear
that the assumption of compact $X$ is only needed to guarantee the
validity of the lemma above. Thus as long as the images by $\mu$ of
the connected components of the critical sets ${\rm Crit} (H_\theta)$
are convex sets, the proposition is still valid, even if $X$ is not
compact. This happens for instance with the linear torus actions on
${\mathbb C}^n$ .
\end{rem}

\end{document}